\documentclass[usenatbib]{mn2e}
\usepackage{epsfig}
\usepackage{graphicx}
\usepackage{amssymb}
\usepackage{aas_macros}
\usepackage{color}

\newcommand{\oversim}[2]{\protect{\mbox{\lower0.5ex\vbox{%
   \baselineskip=0pt\lineskip=0.2ex
   \ialign{$\mathsurround=0pt #1\hfil##\hfil$\crcr#2\crcr\sim\crcr}}}}} 
\newcommand{\simgreat}{\mbox{$\,\mathrel{\mathpalette\oversim>}\,$}} 
\newcommand{\simless} {\mbox{$\,\mathrel{\mathpalette\oversim<}\,$}} 

\begin{document}

\title[Tidal disruption of Globular Clusters in Dwarf Galaxies]{Tidal disruption of globular clusters in dwarf galaxies with triaxial dark matter haloes}

\author[Pe\~{n}arrubia, Walker \& Gilmore]{Jorge Pe\~{n}arrubia$^{1,2}$\thanks{jorpega@ast.cam.ac.uk}, Matthew G. Walker$^{1}$ \&  Gerard Gilmore$^{1}$  \\
$^{1}$Institute of Astronomy, University of Cambridge, Madingley Road, Cambridge, CB3 0HA, UK\\
$^{2}$Department of Physics and Astronomy, University of Victoria, 3800 Finnerty Road, Victoria, BC V8P 5C2, Canada\\
}

 \maketitle  
 
\begin{abstract} 
We use N-body simulations to study the tidal evolution of globular clusters (GCs) in dwarf spheroidal (dSph) galaxies. Our models adopt a cosmologically motivated scenario in which the dSph is approximated by a static NFW halo with a triaxial shape.  We apply our models to five GCs spanning three orders of magnitude in stellar density and two in mass, chosen to represent the properties exhibited by the five GCs of the Fornax dSph.  We show that only the object representing Fornax's least dense GC (F1) can be fully disrupted by Fornax's internal tidal field---the four denser clusters survive even if their orbits decay to the centre of Fornax.  For a large set of orbits and projection angles we examine the spatial and velocity distribution of stellar debris deposited during the complete disruption of an F1-like GC.  Our simulations show that such debris appears as shells, isolated clumps and elongated over-densities at low surface brightness ($\geq 26$ mag/arcsec$^2$), reminiscent of substructure observed in several MW dSphs.  Such features arise from the triaxiality of the galaxy potential and do {\it not} dissolve in time.  The kinematics of the debris depends strongly on the progenitor's orbit.  Debris associated with box and resonant orbits does not display stream motions and may appear ``colder''/``hotter'' than the dSph's field population if the viewing angle is perpendicular/parallel to progenitor's orbital plane. In contrast, debris associated with loop orbits shows a rotational velocity that may be detectable out to few kpc from the galaxy centre.  Chemical tagging that can distinguish GC debris from field stars may reveal whether the merger of GCs contributed to the formation of multiple stellar components observed in dSphs.
\end{abstract} 

%

\section{Introduction}
The sizes ($r_{half} \sim 100$ pc), velocity dispersions ($\sigma_V\sim 10$ km s$^{-1}$) and luminosities ($L_V\sim 10^{3-7} L_{\odot}$) of dwarf spheroidal (dSph) galaxies imply that these objects are the most dark-matter dominated galaxies known (see reviews by Mateo 1998 and Gilmore et al. 2007).  Their dynamical mass-to-light ratios are sufficiently large ($M/L_V\sim 10^{1-3}[M/L_V]_{\odot}$) that their stars can be treated as test particles tracing their underlying gravitational potentials.  Dynamical constraints on the dark matter distribution in dSphs may therefore be compared directly with the predictions from cosmological simulations to study the elusive nature of dark matter (e.g. Gilmore et al. 2007 and references therein).

In the last few years, however, wide-field imaging and multi-object spectroscopic surveys have revealed that dSph stellar components are more complex than previously believed. Localized oddities such as shells (Coleman et al. 2004, 2005; Olszewski et al. 2006), butterfly-shaped substructures (Stetson \& Hesser 1998) and cold clumps (Kleyna et al. 2003, 2004; Ibata et al. 2006)
have been reported in several dSphs. These results conflict with the naive expectation that dSph dark matter haloes (with typical crossing times of a few hundred Myr) should host well-mixed stellar populations and may pose a challenge to the present cosmological paradigm: Cold Dark Matter (CDM). 
In a CDM cosmogony, dark matter haloes follow a density profile that is nearly independent of their mass
(Navarro, Frenk \& White 1996, 1997; hereafter NFW) and scales as $\rho \propto r^{-\gamma}$ in the inner-most regions. Navarro et al. (2008) have recently shown that $\gamma$ varies slightly with halo mass and that dark matter halo profiles only deviate slightly from NFW models ($\gamma=1$).
In such ``cuspy'' potentials, unbound stellar substructures would dissolve after a few crossing times, as shown by the N-body simulations of Kleyna et al. (2003), who also show that the lifetimes of these substructures would be considerably prolonged if the halo has a constant-density ``core''. Thus it has been argued that the presence of stellar substructures in dSphs is in apparent contradiction with CDM expectations.  

A second oddity in this context is the existence of globular clusters (GCs) in dSphs. Although GCs are considerably more common in dwarf irregular galaxies (e.g. Georgiev et al. 2008), at least two of the Milky Way dSphs---Fornax and Sagittarius---contain GCs (five and four, respectively). Among the Milky Way dSphs these galaxies are also the largest and most luminous (Mateo 1998), which may be related to their ability to form GCs at an early epoch.  The extended spatial distribution of Fornax's GCs (most lie, in projection, near Fornax's half-light radius of $\sim 400$ pc) has fueled arguments against cuspy NFW profiles in dSphs.  For example, Goerdt et al. (2006) demonstrate that dynamical friction, proportional in strength to the background density, requires only a few Gyr to drag GCs from Fornax's half-light radius to its centre in a cuspy potential.  Oh, Lin \& Richer (2000) note a possible solution; namely, that strong tidal interactions between the Milky Way and Fornax could inject energy into the GC orbits and delay their decay. However, this scenario is now ruled out by proper motion measurements (Dinescu et al. 2004; Piatek et al. 2007; Walker et al. 2008) that imply Fornax follows a nearly circular orbit around the Milky Way. 
Hernandez \& Gilmore (1998), and also more recently Goerdt et al. (2006) and Read et al. (2006), pose a second viable solution, showing that dynamical friction is suppressed in the inner regions of cored halos, thereby providing a natural explanation for the apparent levity of Fornax's GCs.  The survival of GCs in dwarf galaxies may then provide important clues regarding the inner structure of dark matter haloes.

Despite recent theoretical efforts to understand the formation mechanism of GCs in a cosmological context (e.g. Prieto \& Gnedin 2008, Bournaud et al. 2008), the formation of GCs in dwarf spheroidals lacks definite theoretical underpinning. For example, it is unclear whether Fornax and Sagittarius are the only MW dSphs that satisfy the required conditions to host GCs or whether GCs also formed in smaller and fainter dSphs but did not survive to the present epoch. 
The present contribution 
elaborates on the latter scenario by studying the survival of Globular Clusters in the tidal field of a typical dwarf galaxy and attempts to address the signatures that the tidal disruption of stellar clusters may have left in the stellar populations of dSphs. We examine these issues here with the aid of N-body simulations that adopt a CDM-motivated dSph model, where the dwarf galaxy potential is assumed to follow an NFW profile with triaxial shape. 
This is a timely issue in light of ongoing spectroscopic surveys involving thousands of target stars in  Local Group dSphs.

This paper is organized as follows. Section~2 introduces the observational data and the modelling procedure. Section~3 describes the numerical technique and its limitations. In Section~4 we study the mass and orbital evolution of GCs in NFW halos. In Section~5 we examine the spatial and velocity distribution of disrupted GC debris. Section~6 discusses the possible detection of debris in dSphs. We end with a brief summary in Section~7.

%

\section{Models and Observational data}

Our models assume that the mass profile of a dSph can be approximated by a {\it triaxial} Navarro, Frenk \& White (1996, 1997) dark matter halo. Due to the high mass-to-light ratios estimated in dSphs (e.g. Mateo 1998, Simon \& Geha 2007, Pe\~narrubia, McConnachie \& Navarro 2008) stars are assumed to contribute negligibly to the overall potential. 

Our GC models are N-body realizations of a King (1966) profile orbiting within a static NFW potential. The masses of GCs are relatively small in comparison with the host halo mass and here we assume that the presence of globular clusters (GCs) does not alter the distribution of dark matter.  Dynamical friction resulting from two-body interactions between the GC and the background dark matter particles is introduced in our computations as an external friction term following the method outlined in Pe\~narrubia et al. (2006) (see \S~\ref{ssec:numsetup}). We also assume that the dSph models are in isolation, thereby neglecting any effects that external forces (e.g. the Milky Way tidal field) may induce on the orbital evolution of the GC.  

For illustrative purposes, our study focuses on the dynamical evolution of a GC population chosen to resemble that of the Fornax dSph.  Fornax hosts the largest number of known GCs (five) among all Local Group dSphs.  Insofar as the scatter in parameters characterizing the dark matter halos surrounding MW dSphs is small (e.g., Pe\~narrubia et al. 2008a,b), and given the wide range of masses and spatial sizes covered by Fornax's cluster population, our results should capture the most relevant features of the evolution of GCs in dwarf galaxies.  In the remainder of this Section we provide further details on the numerical methods.

\subsection{The dwarf galaxy model}\label{sec:model}
\label{ssec:dwarfmodel}
We assume that the dSph galaxy is embedded in a static dark matter halo that follows an NFW density profile,
\begin{eqnarray}
\rho_{\rm NFW}=\frac{\rho_0}{(r/r_s)(1 + r/r_s)^2}; \\ \nonumber
\rho_0=\frac{V_{\rm max}^2 r_{\rm max}}{4\pi G r_s^3[\ln(1+r_{\rm max}/r_s)-r_{\rm max}/r_s/(1+r_{\rm max}/r_s)]},
\label{eq:rhonfw}
\end{eqnarray}
where $r_s$ the scale radius and $V_{\rm max}\equiv V_c(r_{\rm max})$ is the peak circular velocity. For NFW halos, $r_{\rm max}\approx 2.17 r_s$.  The halo mass at a given radius is simply $M(r)=4\pi r_s^3\rho_0 [\ln(1+r/r_s)-r/r_s/(1+r/r_s)]$.

In the Cold Dark Matter paradigm, halos generally have a triaxial shape (Davis et al. 1985, Frenk et al. 1998, Jing et al. 1995, Jing \& Suto 2002) that depends only weakly on radius (Kazantzidis et al. 2004b, Allgood et al. 2006).
We account for that effect by introducing elliptical coordinates, using the transformation $r\rightarrow m$, where 
\begin{equation}
m^2=\frac{x^2}{a^2}+\frac{x^2}{b^2}+\frac{z^2}{c^2},
\label{eq:m}
\end{equation}
and $(a,b,c)$ are dimensionless quantities. Our coordinate frame is selected so that $a\ge b\ge c$.  We have used the formulae of Chandrasekhar (1969) to solve the Poisson equations in elliptical coordinates for the density profile of eq.(\ref{eq:rhonfw}). The resulting potential can be written as
\begin{eqnarray}
\Phi_{\rm NFW}(x,y,z)=2\pi G a b c \rho_0 r_s^2 \\ \nonumber
\times \int_0^\infty \frac{m(\tau)/r_s}{1+m(\tau)/r_s}\frac{d\tau}{\sqrt{a^2+\tau}\sqrt{b^2+\tau}\sqrt{c^2+\tau}} + {\rm const},
\label{eq:potnfw}
\end{eqnarray}
where $m^2(\tau)\equiv x^2/(a^2+\tau) + y^2/(b^2+\tau)+ z^2/(c^2+\tau)$. We choose a triaxiality of $(a,b,c)=(1.47,1.22,0.74)$, which roughly corresponds to the most common shape found in cosmological simulations (e.g. Jing \& Suto 2002). 
The recent work of Allgood et al. (2006) shows that haloes tend to become slightly rounder as their mass decreases and also at outer radii. Because GCs in dSphs are only found in the inner-most regions of the dwarf dark matter haloes (see below), the triaxiality gradients found by Allgood et al. (2006) can be safely neglected for this study. Note also that choosing a different combination of axis-ratios will change the phase-space volume occupied by each orbital family (see e.g. Fig.~1 of Adams et al. 2007) without affecting the qualitative analysis on the dynamical evolution of GCs in a triaxial potential outlined in the following Sections. 

The two free parameters of the NFW potential ($V_{\rm max}$ and $r_{\rm max}$) are scaled to the values estimated for MW dSphs by Pe\~narrubia et al. (2008a,b) who find a remarkably small scatter in dSph masses. Indeed, the studied dSphs differ by less than a factor $\sim 4$ in $V_{\rm max}$, with peak velocities in the range $10 \le V_{\rm max} \le 40$ km/s and peak radii $2\simless r_{\rm max} \simless 9$ kpc, despite the fact that their luminous components span four decades in luminosity ($\sim 10^3$--$10^7 L_\odot$) and one decade in spatial size (100--1500 pc core radii).  Systems of particular interest for this contribution are Sagittarius and Fornax, as these are the only MW dSphs with associated populations of gravitationally bound GCs. The estimated NFW parameters of these galaxies are, respectively, $V_{\rm max}=13.8$ and 20.6 km/s, with peak velocity radii at $r_{\rm max}=2.6$ kpc and 4.1 kpc. 

Finally, we note that in our dSph models stars and GCs are confined to the inner-most regions of the halo (i.e. the ``cusp'' in an NFW halo, see Pe\~narrubia et al. 2008a). Due to (i) the small amount of dark matter substructures in subhaloes with respect to that in the main halo and (ii) the large galactocentric distances at which substructures tend to locate (Springel et al. 2008), for simplicity we neglect here the effects that the halo granity may induce on the orbits of cluster particles and assume a smooth potential for our dwarf halo models.

\subsection{Globular Cluster models}
\label{ssec:gcmodel}

Our N-body realizations of GCs assume a King (1962) stellar density distribution,
\begin{eqnarray}
\rho_\star=\frac{K}{x^2}\
\bigg[\frac{\cos^{-1}(x)}{x}-\sqrt{1-x^2}\bigg],~~x\equiv\bigg[\frac{1+(r/r_c)^2}{1+(r_t/r_c)^2}\bigg]^{1/2},
\label{eq:rhok}
\end{eqnarray}
where $r_c$ and $r_t$ are the King and tidal radii, and $K$ is an arbitrary constant that we select as $K=L / \int_0^{r_t} \rho_\star d^3 r$, where $L$ is the total luminosity of the cluster. All our cluster models have a particle number $N=10^5$.
 
For comparison with observations it is easier to use the projected surface density
\begin{eqnarray} 
\Sigma (R)= k \Bigg\{  \frac{1}{ \big[1+(R/R_c)^2\big]^{1/2} }-\frac{1}{ \big[1+(R_t/R_c)^2\big]^{1/2} }  \Bigg\}^2,
\label{eq:Sigma} 
\end{eqnarray} 
where $R$ is the projected radius, $k$ is related to the central surface brightness by $\Sigma_0=k\big(1- 1/\sqrt{1+c_K^2}\big)^2$ and  $c_K\equiv R_t/R_c$. 
 
Table~\ref{tab:obs} lists the properties of the GC population of the Fornax dwarf relevant for this study. For comparison we also include the properties of the four clusters in the Sagittarius dwarf.  Throughout this work we convert luminosities into masses and {\it vice versa} assuming GCs have stellar mass-to-light ratios of approximately $(M/L)_\star\approx 3$ (e.g. Bruzual \& Charlot 1993, Bell \& de Jong 2001).  As we show below, the orbital and dynamical evolution of a GC due to dynamical friction and tidal disruption depend primarily on the GC's mass and density.

\begin{figure}
  \includegraphics[width=84mm]{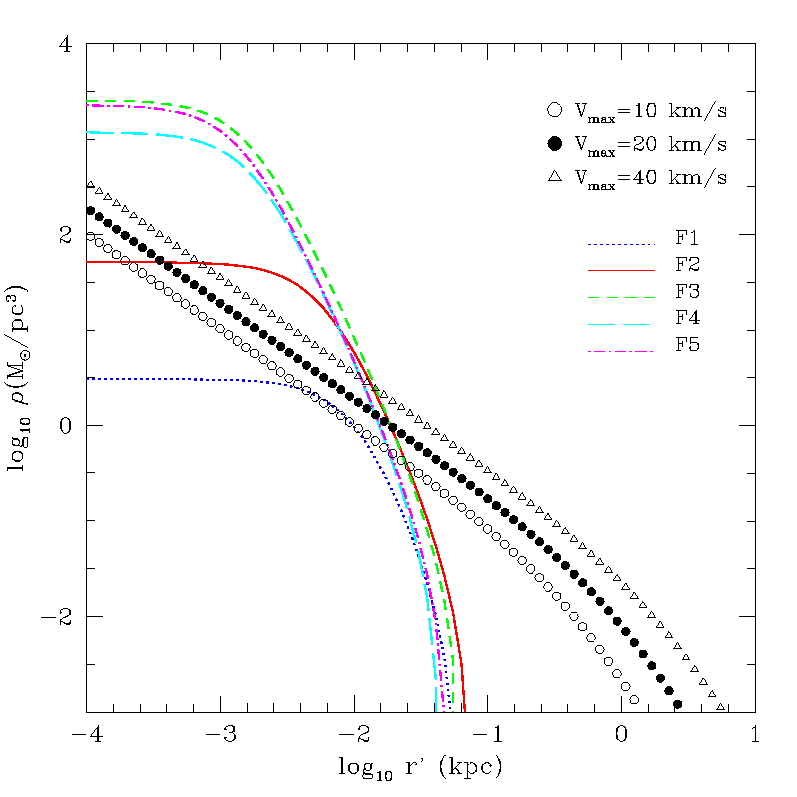}
\caption{Density profiles of the five GCs in Fornax (F1,...,F5) and of the underlying dark matter halo. The halo of the Fornax dSph is assumed to follow an NFW profile with $V_{\rm max}\simeq 20$ km/s. For comparison we also plot the density profiles of halos with $V_{\rm max}=10$ km/s and 40 km/s, which represent the approximate range of maximum circular velocities of the Local Group dSphs (Pe\~narrubia et al. 2008b). We distinguish three categories of clusters: Those with central densities higher (F3, F4 and F5), comparable (F2) and lower (F1) than that of the dark matter halo at $r\sim R_c$.}
\label{fig:dens}
\end{figure}

Fig.~\ref{fig:dens} compares the density profiles of the Fornax GCs to those of the NFW halos for the whole range of $V_{\rm max}$ values estimated for the Milky Way dSphs by Pe\~narrubia et al. (2008a).  In particular, filled circles show the density profile of the Fornax halo.  This figure demonstrates two relevant points.  First, Fornax's GCs divide into three categories based on their central densities: (i) low-density clusters like F1, with $\rho_\star(0) \simless \rho_{\rm NFW}(R_c)$; (ii) intermediate-density clusters (F2), $\rho_\star(0) \simgreat \rho_{\rm NFW}(R_c)$ and  (iii) high-density clusters like F3, F4 and F5, with $\rho_\star(0)\gg \rho_{\rm NFW}(R_c)$. Henceforth, to simplify our study we will consider only the clusters F1, F2 and F3 to represent the three GC categories.
Second, the NFW halos estimated for the MW dSphs display a small scatter in density at small radii ($r\ll r_s$). Varying the peak velocity between 10 and 40 km/s leads to a density variation of a factor $\simless 3$ at a given radius (e.g. at 100 pc), whereas the stellar densities of the Fornax and Sagittarius GCs span a factor $\sim 1000$. Then the domination of dSph gravitational potentials by similar dark matter halos implies that the evolution of GCs depends primarily on the clusters' properties at the time of formation, rather than on the properties of their host halos.

The second parameter relevant to this study is the mass of the GC. Taking into account that all the bright dwarfs in the Milky Way appear to have dynamical masses close to $M_{\rm dyn}\sim 10^7 M_\odot$ (e.g. Mateo 1998), GCs represent only a small fraction of the dwarf mass, with mass ratios in the range  $M_{GC}/M_{\rm dyn}\sim 10^{-3}$ (F1) and $10^{-2}$ (F3). 

\section{Numerical setup}
\label{ssec:numsetup}
\subsection{The N-body code}

We follow the evolution of the cluster N-body models in the dwarf potential using {\sc Superbox}, a highly efficient particle-mesh gravity code (see Fellhauer et al. 2000 for details). {\sc Superbox} uses a combination of different spatial grids in order to enhance the numerical resolution of the calculation in the regions of interest.  In our case, {\sc Superbox} uses three nested grid zones centered on the highest-density particle cell of the cluster.  This centre is updated at every time step, so that all grids follow the cluster along its orbit.  
  
Each grid has $64^3$ cubic cells: (i) the inner grid has a spacing of $dx_1=10 R_c / 62\simeq 0.16 R_c$ and is meant to resolve the entire Globular Cluster. (ii) The middle grid extends to intermediate distances with a resolution of $dx_2=100 R_c/62 \simeq 1.6 R_c$. (iii) The outermost grid extends out to 250 kpc and is meant to follow particles that are stripped from the cluster and subsequently orbit within the main galaxy.  

In addition to the overall dSph potential, GCs feel dynamical friction as they orbit within the dSph's extended halo. To describe dynamical friction in aspherical haloes we apply the equations of Binney (1977), which recover the Chandrasekahr (1943) formula for $a=b=c=1$. 
Pe\~narrubia, Just \& Kroupa (2004) test the accuracy of these equations against self-consistent N-body simulations and find that, for a large range of orbital parameters and masses, the overall radial evolution of point-masses within flattened dark matter haloes can be reproduced to within a few percent. As the N-body simulations of Pe\~narrubia et al. (2002) show, halo asphericity introduces a strong dependence between the orbital decay time and the orientation of the orbit with respect to the halo's plane of symmetry. In particular, these authors show that orbits that lie on the major--intermediate axis plane have shorter decay times.

For simplicity we assume here that the velocity distribution of the dark matter halo is Maxwellian and isotropic. Kazantzidis et al. (2004a) have examined the inadequacies of the local Maxwellian approximation when applied to NFW haloes and, although it may introduce problems to generate haloes in perfect equilibrium, the deviations from Gaussianity of the exact distribution function have a negligible impact on the friction term (see their Fig. 4).
Under these approximations, the friction force per unit mass can be written as (Binney 1977)
\begin{eqnarray}
{\bf f}_{{\rm df}}=-2\sqrt{2\pi}G\rho_{\rm NFW} \frac{{\bf v}}{\sigma^3} B M_{\rm GC}\ln\Lambda ,
\label{eq:fric}
\end{eqnarray}
where $\sigma^2( m)=\rho^{-1}_{\rm NFW}\int_m^\infty dr' M(r')\rho_{\rm NFW}(r')/r^{'2}$ is the halo velocity dispersion, $B=\int_0^\infty d\tau \exp[-v^2/(1+\tau)/2\sigma^2]/(1+\tau)^{5/2}$ and $M_{\rm GC}$ is the cluster's bound mass.  Pe\~narrubia et al. (2004) find that the best fit to self-consistent N-body orbits is obtained for a Coulomb logarithm $\ln\Lambda=2.1$, in good agreement with the recent results of Arena \& Bertin (2007). However, it is well known that this quantity is fairly sensitive to numerical aspects, such as particle number and spatial resolution (e.g. Prugniel \& Combes 1992, Wahde \& Donner 1996) as well as to the host's density profile (Hashimoto et al. 2003, Just \& Pe\~narrubia 2005, Read et al. 2006, Arena \& Bertin 2007). It is therefore necessary to note that the decay times of clusters orbiting in dark matter haloes are uncertain to a significant degree, since orbital decay rates are proportional to the value assumed for $\ln \Lambda$.

In order to implement eq.~(\ref{eq:fric}), we assume that only particles bound to the Globular Cluster feel dynamical friction, whereas unbound (stripped) particles do not. In this way we neglect effects that the density wake self-gravity may induce on the trajectories of escaping particles. 

After calculating the different force terms, {\sc Superbox} solves the equation of motion of each cluster particle
\begin{eqnarray}
\frac{d^2{\bf r}_i}{dt^2}=-{\bf \nabla}\Phi_{\rm NFW}({\bf r}_i) + \xi {\bf f}_{{\rm df}}({\bf r}_i),
\label{eq:eqmot}
\end{eqnarray}
where $\xi=1$ for bound particles and $\xi=0$ for unbound ones.
  
{\sc Superbox} uses a leap-frog scheme with a constant time-step to integrate the equations of motion for each particle.  Each N-body realization is evolved a Hubble time ($t_H=$14 Gyr) in the NFW dwarf potential. We choose a time step $dt=t_{\rm cr}/25$, where $t_{\rm cr}$ is the cluster crossing time defined as
 \begin{eqnarray}
t_{\rm cr}\equiv 2\pi \frac{R_c^{3/2}}{\sqrt{GM_{GC}}}.
\label{eq:tcr}
\end{eqnarray}
The crossing times of F1, F2 and F3 are, respectively, 15.7, 3.1 and 0.32 Myr, spanning a factor $t_{\rm cr, F1}/ t_{\rm cr, F3}\simeq 50$. Note that the crossing time of the Fornax dwarf is $t_{\rm cr}\simeq 230$ Myr assuming a dynamical mass of $10^7 M_\odot$ and $R_c=400$ pc, illustrating the large range of time-scales that our simulations must cover, since clusters evolve a factor $10$--$1000$ faster than the host galaxy.  The total number of time-steps required to simulate each of the clusters for a Hubble time is therefore large, $1.5\times 10^{4}$, $8.0\times 10^{4}$ and $7.5\times 10^{5}$ for the clusters F1, F2 and F3, respectively.  We find that our choice of numerical parameters leads to a total energy conservation better than 5\% after the cluster models are evolved for a Hubble time in isolation, which yields a negligible evolution of the initial model parameters.

\subsection{Relaxation processes in clusters}\label{sec:rel}
{\sc Superbox} is a collision-less N-body code and therefore does not account for a number of internal processes that take place in clusters, such as mass loss due to stellar evolution, core collapse and the subsequent re-expansion due to two-body relaxation processes. Whereas the evolution associated with stellar evolution is significant only during the early life of a cluster ($\simless 1$ Gyr after formation, Baumgardt \& Makino 2003), two-(and higher order)-body interactions between stars may affect clusters during their entire evolution.  A realistic incorporation of these effects into dynamical simulations has proven to be a theoretical as well as a numerical challenge due to the complexity of the problem and the large range of time-scales involved (e.g. Aarseth \& Heggie 1998). Only in the last few years has the availability of {\sc GRAPE} special-purpose supercomputers provided the necessary computational tools to simulate the evolution of GCs with a number of particles $N\simgreat 10^5$ (e.g. Baumgardt \& Makino 2003).

To estimate the importance of two-body relaxation processes in the Fornax clusters we calculate first the relaxation time as (Binney \& Tremaine 2008, eq.~1.38)
\begin{eqnarray}
t_{\rm rel}\approx \frac{N_\star}{ 8\ln (N_\star\gamma)} t_{\rm cr},
\label{eq:trel}
\end{eqnarray}
where $N_\star\approx M_{GC}/m$ and $m$ is the mean stellar mass in a cluster. A Kroupa (2001) mass function then implies $m\simeq 0.58 M_\odot$. 

Two-body interactions increase the strength of dynamical friction. As a result, the values of the Coulomb logarithm in multi-mass clusters are higher than those found in collision-less systems. The factor $\gamma$ in the Coulomb logarithm depends on the mass spectrum and is rather uncertain for clusters. As an additional complication, the mass spectrum changes during the dynamical evolution of a globular cluster (Baumgardt \& Makino 2003). Here we adopt the value found by Giersz \& Heggie (1996), $\gamma=0.02$.  For this choice of parameters, we estimate that the Fornax clusters F1, F2 and F3 have relaxation times of $t_{\rm rel}\simeq 14.6, 4.4$ and 2.2 Gyr, respectively. Thus, only F1 has a relaxation time $t_{\rm rel}\simgreat t_H$, which suggests that stellar encounters have played only a small role over its lifetime. Fornax's other clusters, however, have likely suffered repeated episodes of core collapse and continuous mass dissolution (H\'enon 1960). In practice, this result implies that their present properties have changed significantly since their formation and cannot be extrapolated to several Gyr in the past.

\section{Mass and orbital evolution of clusters in dwarf galaxies}
Two parameters govern the dynamical evolution of a gravitationally bound system within a host galaxy: (i) the mean density, which regulates the strength of external tides and (ii) the cluster mass, which determines the amount of dynamical friction.
\subsection{Mass vs density}\label{sec:massdens}
\begin{figure}
\includegraphics[width=84mm]{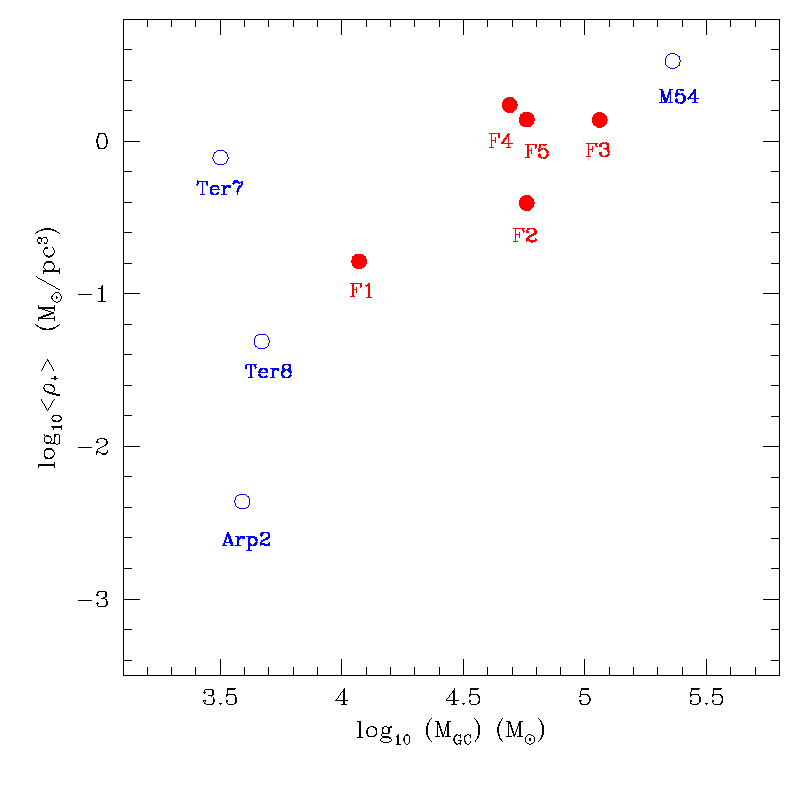}
\caption{ Mean density $\langle \rho_\star \rangle\equiv M_{GC}/(4\pi R_t^3/3)$ versus stellar mass for the GC population in the Fornax (filled symbols) and Sagittarius (open symbols) dSphs. Massive, low-density clusters, which experience the strongest dynamical friction and tidal mass stripping, are absent in dwarf galaxies.}
\label{fig:densmass}
\end{figure}

In Fig.~\ref{fig:densmass} we plot cluster mass against the mean density for the nine GCs of Fornax and Sagittarius.  The large range of properties exhibited by these clusters suggests that tidal forces and dynamical friction act with different strengths. In particular, low-mass, low-density clusters, such as Arp 2, Ter 8 and F1, will suffer a small or negligible friction force, but will respond heavily to tidal interactions with the host galaxy.  In contrast, owing to its high density and low mass, Ter 7 will likely experience little dynamical evolution. At high mass and high density, clusters F3, F4, F5 and M54 may suffer a strong orbital decay as a result of dynamical friction. However, since these clusters are considerably more dense than the underlying dark matter halo (see Fig~\ref{fig:dens}), they should be resilient to tidal mass stripping.

\begin{figure}
\includegraphics[width=84mm]{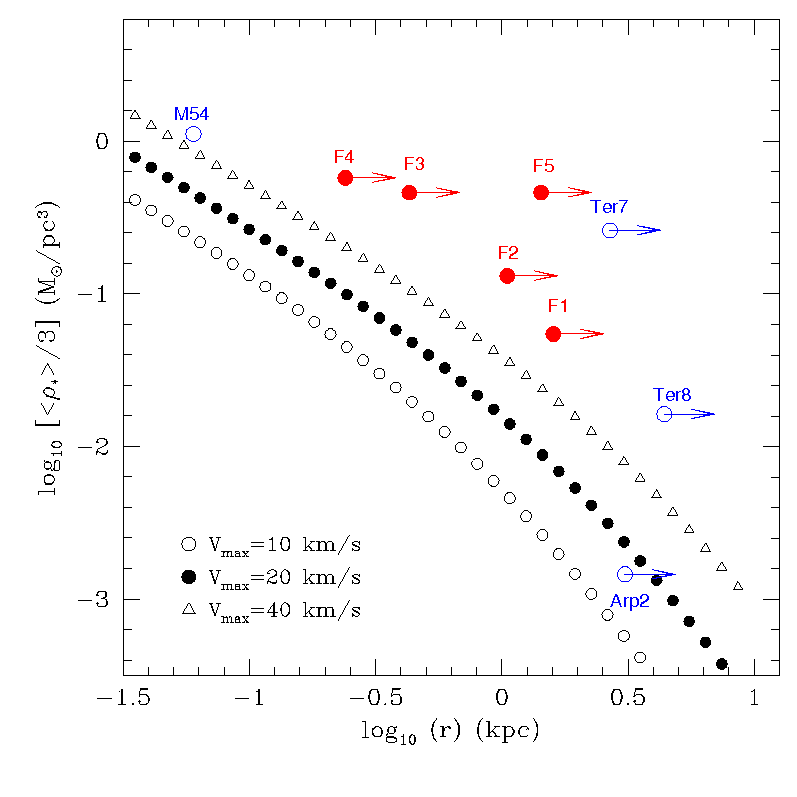}
\caption{Mean stellar density of clusters $\langle \rho_\star \rangle /3$ as a function of their {\it projected} position within the dwarf. Arrows indicate that the projected position is always smaller than or equal to the distance to the dwarf's centre. For comparison, we also plot the mean density of the host's halo $\langle \rho_{\rm NFW}\rangle(r)$ for the range of maximum velocities estimated for the Milky Way dwarfs. Note that the densest and most massive clusters tend to be located in the inner regions of dwarfs, suggesting that the cluster population may have experienced significant dynamical evolution. Also note that the fact that $\langle \rho_\star \rangle /3 > \langle \rho_{\rm NFW}\rangle $ suggests that tidal mass loss no longer affects these systems at their present location.}
\label{fig:denstidal}
\end{figure}

\begin{figure}
\includegraphics[width=84mm]{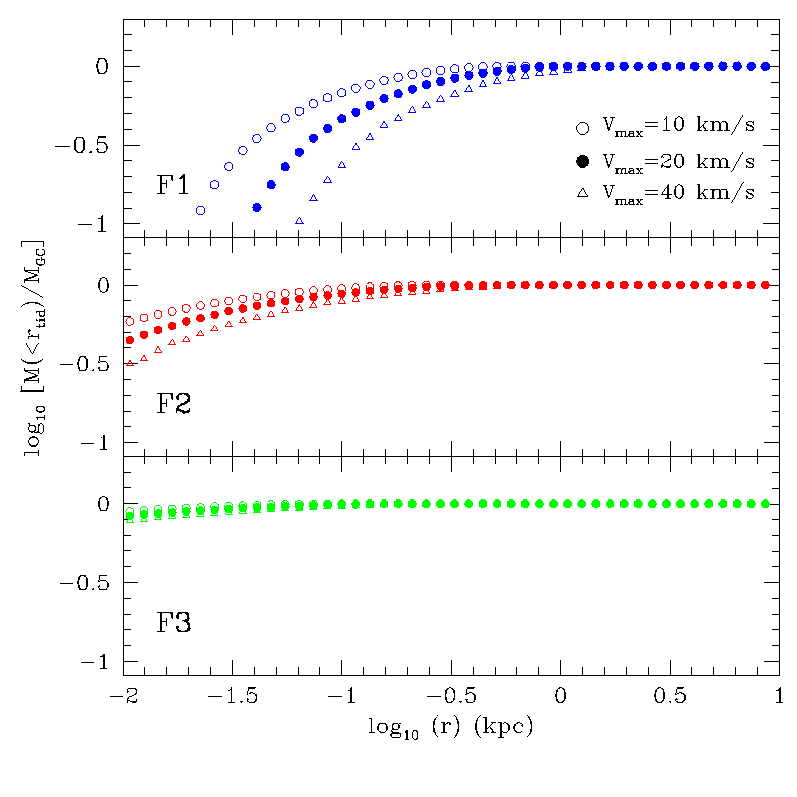}
\caption{Mass fraction enclosed within the tidal radius of the clusters F1, F2 and F3 as a function of the distance to the host centre. Note that the cluster F3 (also F4 and F5) is so dense that it would survive tidal disruption in the potential of any known Milky Way dSph.}
\label{fig:mrtid}
\end{figure}

Despite the fact that the low-mass end of the Fornax/Sagittarius GC distribution shows a remarkable dispersion in stellar densities ($\sim 2.5$ orders of magnitude), no massive, low-density clusters are observed in these dSphs. The absence of massive, low-density GCs may reflect processes relevant to GC formation or, alternatively, may indicate that such clusters have been disrupted by tides as dynamical friction brought them close to the inner-most regions of their host galaxies.  If the latter scenario is correct, the effects of dynamical friction and tidal stripping should give rise to spatial segregation according to GC mass and density.  Specifically we expect surviving low-mass GCs to populate the outskirts of their host halos, and we expect the most massive GCs to have been dragged to the innermost regions, where they can survive only if sufficiently dense.

\subsection{Density vs projected position}
Indeed, Fig.~\ref{fig:denstidal} shows that the Globular Clusters of Fornax and Sagittarius exhibit a clear spatial segregation.  Dots show the estimated density of the underlying dark matter halos and arrows indicate that the projected radius is the minimum possible distance to the dwarf centre\footnote{M54 sits in the centre of the Sgr dwarf. To plot this cluster in Fig.~\ref{fig:denstidal} we adopt a galactocentric distance equal to its King tidal radius}, i.e $R\le r$. As expected, Arp 2, Ter 8 and F1 --- the lowest-density clusters in our sample--- are found farthest from the dwarf centre, whereas M54 and F3 --- the densest and most massive clusters--- reside in the inner regions.  

A second interesting result is that GCs orbiting dwarf galaxies tend to reside at a projected radius, $R$, such that the mean density of the cluster $\langle \rho_\star\rangle\equiv M_{GC}/(4\pi R_t^3/3)$ is $\langle \rho_\star\rangle /3\simgreat \langle \rho_{\rm NFW}\rangle (R)$, which suggests that tidal mass stripping no longer affects the existing GCs at their present locations (see \S~\ref{sec:tidal}).  As in the case for the mass-density trend shown in Fig.~\ref{fig:densmass}, it is unclear whether the correlation between cluster density and position apparent in Fig.~\ref{fig:denstidal} arises from a formation mechanism or results from the disruption of low-density systems.  In this contribution we focus on the latter scenario. We begin by examining the effect of dynamical processes on the evolution of GCs in a dwarf galaxy potential.

\subsection{How resilient are GCs to tidal forces?}\label{sec:tidal}
We quantify the expected influence of tides by computing the theoretical ``tidal radius'' of the cluster, $r_{\rm tid}$, as 
\begin{equation}
\langle \rho_\star\rangle (r_{\rm tid})/3=\langle \rho_{\rm NFW}\rangle (r),
\label{eq:rtid}
\end{equation}
 where $\langle \rho_{\rm NFW}\rangle$ denotes the mean enclosed density within the cluster's galactocentric radius $r$. The estimated amount of bound mass is $M_{GC}\equiv M_{GC}(<r_{\rm tid})$. In Fig.~\ref{fig:mrtid} we plot this quantity for the clusters F1, F2 and F3 as a function of their distance to the Fornax centre. Our simple analytic estimates provide a simple means of predicting the dynamical evolution of GCs.  First, the cluster F3 (and similarly F4 and F5) is sufficiently dense (see Fig.~\ref{fig:dens}) that it should withstand the tidal field of Fornax even in the inner-most regions of the galaxy. In fact, since the Milky Way dwarf spheroidals have peak velocities $V_{\rm max}\simless 40$ km/s (Pe\~narrubia et al. 2008a,b), our analytical estimates indicate that, if two-body relaxation processes are ignored, the clusters F3, F4 and F5 would likely survive tidal disruption in any of those galaxies.  Second, intermediate-density clusters, such as F2, will suffer tidal mass loss if their orbits bring them to galactocentric distances similar to their spatial size ($r\simless R_{t, GC}$). Finally, these estimates show that low-density clusters like F1 may suffer strong episodes of tidal mass stripping (and eventually, disruption) for a large distance range. Stellar debris resulting from the cluster disruption will dissolve in the dwarf galaxy and may be detectable with the present instrumentation. We shall return to this point below.

\begin{figure}
\includegraphics[width=84mm]{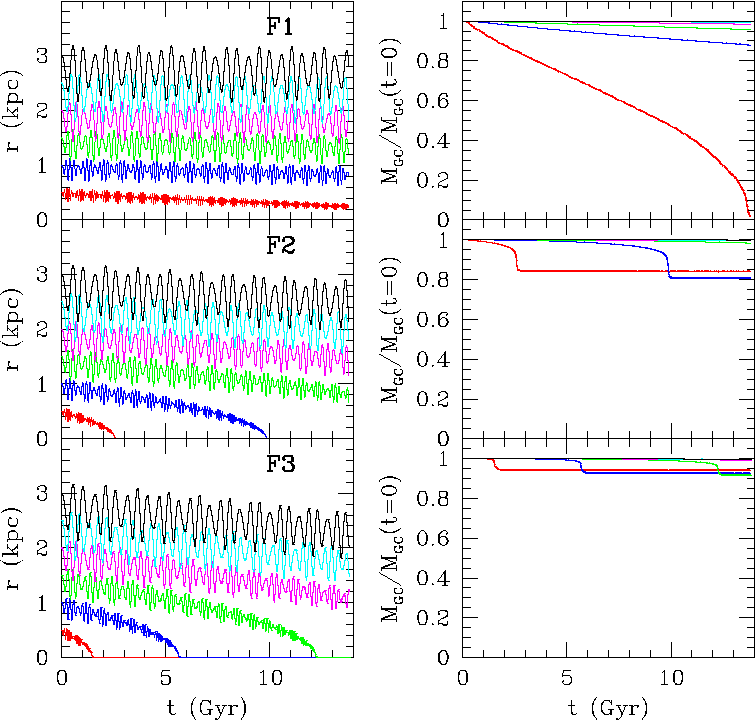}
\caption{ Galactocentric distance ({\it left column}) and mass ({\it right column}) evolution of the clusters F1, F2 and F3 in a dwarf galaxy halo with a peak velocity $V_{\rm max}=20$ km/s.  Different colours denote different initial galactocentric distances ($r_0$). The orbits are confined to the $X-Y$ plane (i.e., the plane formed by the halos's long and intermediate axis) and have an orbital circularity $\eta=1$ (i.e. the initial velocity $v_0$ is such that $v_0(r_0)=V_c(r_0)$).}
\label{fig:rmevol}
\end{figure}

\subsection{Dynamical evolution of clusters}\label{sec:dynev}
To examine the analytical results outlined in \S\ref{sec:tidal} we have carried out a simple numerical experiment. Fig.~\ref{fig:rmevol} shows the evolution of the galactocentric distance (left panels) and bound mass fraction (right panels) for N-body realizations of the clusters F1, F2 and F3 in a dwarf galaxy potential that has parameters similar to those of Fornax (see \S\ref{ssec:dwarfmodel}).  For simplicity, we consider orbits confined to the X-Y plane (i.e. the plane formed by the long and intermediate axis of the dwarf's halo, respectively).  We place model GCs at different galactocentric distances $r(t=0)\equiv r_0$ with  an initial velocity that corresponds to the circular velocity of the Fornax dwarf at the cluster's initial location, i.e $v_0=V_c(r_0)$. The oscillation of the orbital radius around its initial value, $r_0$, results from the triaxiality of the halo potential (see e.g. \S~3.3 of Binney \& Tremaine 2008).

As expected from the estimates shown in Fig.~\ref{fig:mrtid}, tidal stripping has a minor impact on the evolution of the clusters F2 and F3. Neglecting the internal evolution induced by relaxation processes in these systems, our simulations show that the tidal field of Fornax can strip a maximum 20\% and 10\% of the initial mass of the clusters F2 and F3, respectively. 

In contrast dynamical friction induces a strong orbital decay of F2 and F3 {\it only} for those orbits that are initially placed within $r_0\simless 1.5$ kpc from the dwarf galaxy centre. The orbital decay process is further strengthened by the small amount of mass that these dense clusters lose. 
The effects of dynamical friction become considerably weaker as we increase $r_0$. F2 and F3 models that are initially placed beyond  $\simgreat 1.5$ kpc do not sink to the galaxy centre in a Hubble time. 
These results are in good agreement with those of Goerdt et al. (2006), considering that these authors use a different numerical setup. Their models assume point-mass clusters orbiting in a live, self-consistent NFW halo with parameters similar to those adopted here. The fact that we derive similar decay times lends support to the semi-analytical treatment of dynamical friction presented in \S\ref{ssec:numsetup}.  

Although the present distances of the clusters F2, F3, F4 and F5 to the centre of Fornax are unknown, their projected locations fall within the dwarf's limiting radius (see Table~\ref{tab:obs}). If we adopt the {\it projected} position as the galactocentric distance, the fact that these clusters did not sink to the galaxy centre within a Hubble time suggests that they formed in the outer regions of Fornax and that dynamical friction has had little influence on their orbital evolution. This explanation, however, places strong constraints on theoretical models of the formation of GCs in dwarf galaxies, a topic that still lacks theoretical underpinning.

We note here that Hernandez \& Gilmore (1998) and Goerdt et al. (2006) propose an alternative explanation for the non-central positions of the Fornax GCs. These authors show that adopting a {\it cored} halo profile can efficiently stall the orbital decay process. Numerical simulations show that in cored haloes dynamical friction is suppressed within the halo core (Read et al. 2006) and that, as a result, the orbital decay of GCs stalls at the core radius and does {\it not} proceed to the galaxy centre.   However, M54 presents a challenge for this solution. M54 is the densest and most massive cluster found in the Milky Way dSph population and presently sits in the centre of the Sagittarius dSph. Its chemical composition and kinematics are clearly distinct from those of the surrounding stars (Bellazzini et al. 2008), which indicates that M54 formed elsewhere and was dragged to its present location by dynamical friction, in a fashion similar to our models shown here.

Finally, Fig.~\ref{fig:mrtid} shows that the cluster F1 suffers a negligible orbital decay regardless of the initial galactocentric radius $r_0$.  Tidal mass stripping has negligible impact on the models that orbit beyond $\simgreat 0.5$ kpc. Smaller galactocentric distances, however, lead to the complete tidal disruption of F1 within a Hubble time.  In the following Sections, we use F1 as a testbed to study the observational imprints that the tidal disruption of globular clusters may have left on dSph stellar populations.

\begin{figure}
\includegraphics[width=84mm]{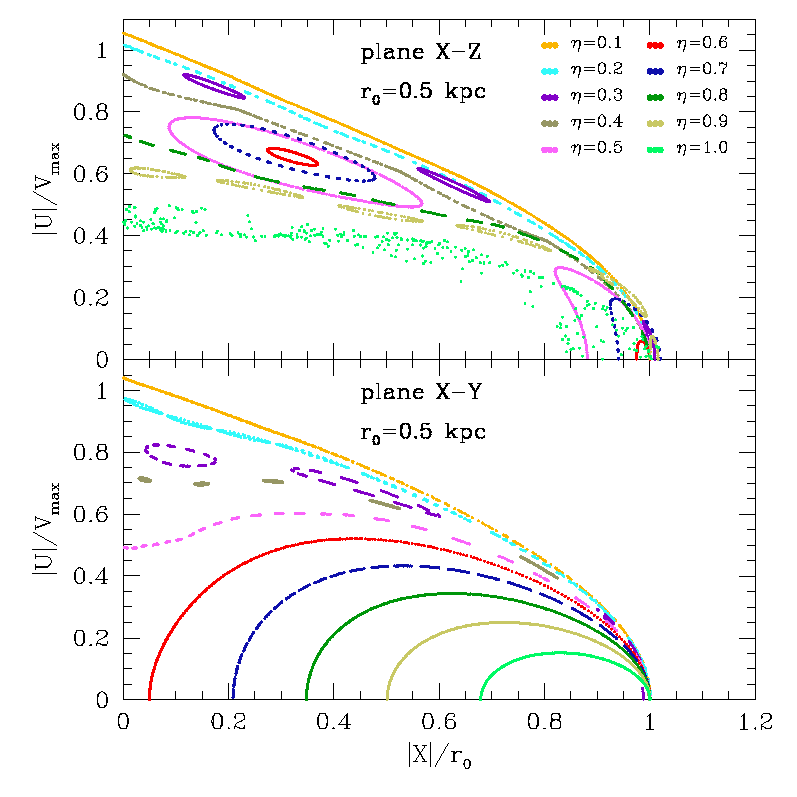}
\caption{ $(X,U)$ surface of section of the orbits placed at an initial distance $r_0=0.5$ kpc from the dwarf centre. The orbital planes are defined by the major and intermediate axis ({\it lower panel}) and by the major and minor axis ({\it top panel}). The parameters of our dwarf galaxy halo have been scaled to the values estimated for the Fornax dSph, $V_{\rm max}=20.6$ km/s and $r_{\rm max}=4.1$ kpc.}
\label{fig:phase}
\end{figure}

\begin{figure}
\includegraphics[width=84mm]{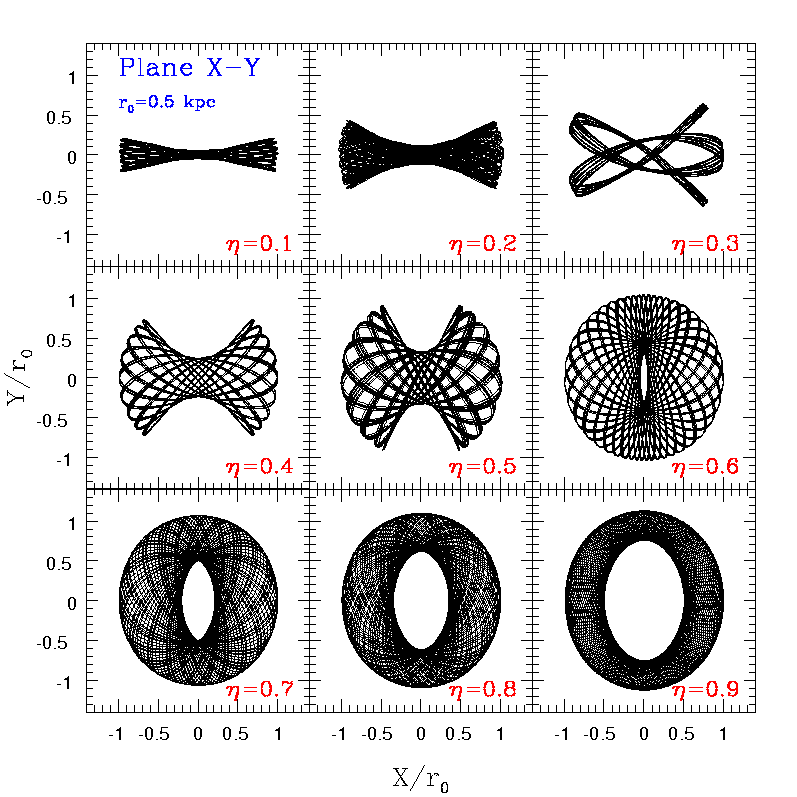}
\includegraphics[width=84mm]{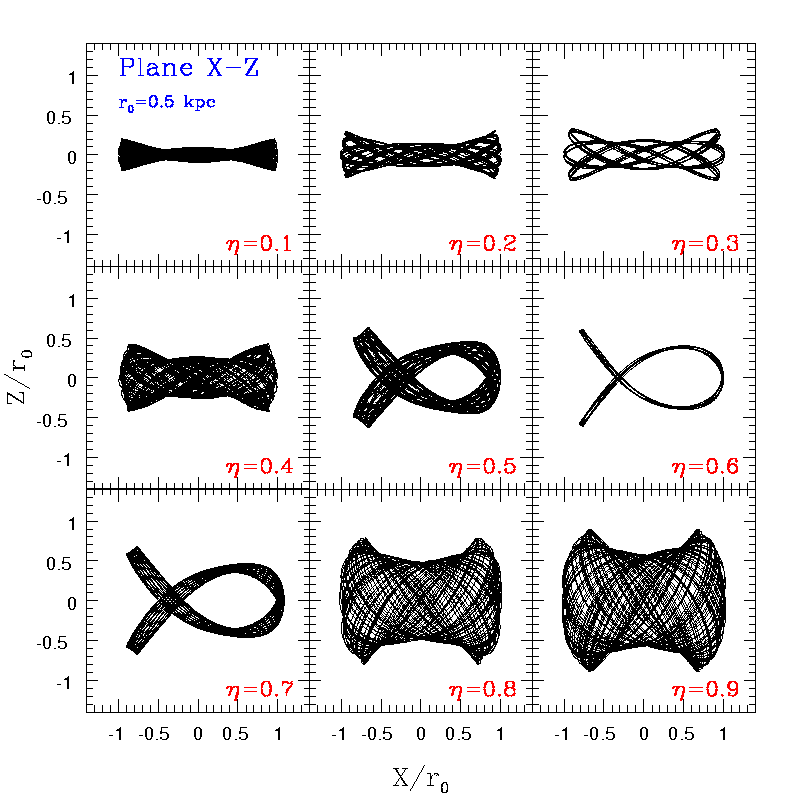}
\caption{Orbits on the $X-Y$ (upper panel) $X-Z$ (lower panel) planes with orbital apocentre $r_0=0.5$ kpc for different values of circularity $\eta$. }
\label{fig:orbXY}
\end{figure}


\section{Disruption of clusters in triaxial potentials}\label{sec:disrupt}
 
\subsection{The orbits}\label{sec:orbits}
According to CDM modelling, dwarf galaxies are embedded in dark matter halos that have a triaxial shape. Orbits in triaxial potentials are usually classified in four major families: (i) boxes, (ii) loops (also called tubes), (iii) resonant orbits and (iv) irregular (also called unstable or stochastic) orbits. The fraction of phase-space volume occupied by each of those families depends on several factors, for example mass profile and shape of the galaxy (e.g. Schwarzschild 1979, 1982; de Zeeuw \& Merritt 1983; de Zeeuw 1985; Miralda-Escud\'e \& Schwarzschild 1989; Merritt \& Valluri 1999). The study of orbits in triaxial potentials is a complex topic and beyond the scope of this contribution. We refer interested readers to \S~3.3 of Binney \& Tremaine (2008) and the papers listed above for more detailed analysis.

For simplicity, we restrict our study to orbits that initially lie on the $X-Y$ and $X-Z$ symmetry planes; recall that the coordinates $(X,Y,Z)$ are oriented along the major, intermediate and minor axis, respectively. Our N-body models are placed along the major axis at an initial distance $r_0$. The range of initial distances explored here, $r_0\in[0,3]$ kpc, extends considerably farther than the stellar truncation radius of Fornax ($R_{t,{\rm For}}\approx 2.1$ kpc). To select the initial velocity, $v_0$, we define the {\it orbital circularity} $\eta\equiv J/J_c(E)$, where $J=r_0 v_0$ is the angular momentum and $J_c(E)$ the angular momentum of a circular orbit with the same energy $E=1/2 v_0^2+ \Phi_{\rm NFW}(r_0,0,0)$. This parameter has values in the range $\eta\in[0,1]$. Because in triaxial systems the angular momentum is not a constant of motion, $\eta$ 
is only well defined at $t=0$.  However, introducing this parameter to define our initial set of orbits allow us: (i) to vary the initial orbital energy at a fixed orbital apocentre in a simple way and (ii) to facilitate a comparison with models in spherical potentials in order to address the effects of triaxiality (see \S\ref{sec:triaxeff}).

Fig.~\ref{fig:phase} shows the surface of section of the orbits placed initially at $r_0=0.5$ kpc. In this plot, each dot corresponds to a crossing of the plane specified by $Z=Y=0$. The first result that stands out is the chaotic nature of the orbits, in the sense that small variations in the initial velocity/circularity can lead to completely different orbit configurations. This plot includes examples of the four orbit families mentioned above
\begin{itemize}
\item {\bf Box orbits}: Systems on box orbits repeatedly pass close to the centre of the potential, and eventually at arbitrarily close distances. They have no net angular momentum and therefore no sense of rotation. Examples are the orbits with $\eta \le 0.2$ in both panels of Fig.~\ref{fig:phase}.

\item {\bf Resonant orbits}: Resonant orbits satisfy a relation of the form $l w_X+ m w_Y + n w_Z=0$, where $w_i$ ($i=X,Y,Z$) are the fundamental frequencies around the principal symmetry axis and $l,m,n$ integers, not all of which are zero. In the surface of section plotted in Fig.~\ref{fig:phase}, resonant orbits can be seen as chains of islands (for example, orbits with $\eta=0.3$ in both panels. Also orbits with $\eta=0.5, 0.6$ and 0.7 moving on the $X-Z$ plane are clear examples of resonant orbits). An aspect of these orbits that will prove crucial for our study is that, in contrast to boxes, resonant orbits are {\it centrophobic}, i.e., they avoid the centre of the potential.

\item {\bf Loop orbits}: Systems moving on loops have a definite sense of rotation about the galaxy centre and also avoid the region of the origin. Interestingly, we find loop orbits only on the plane $X-Y$. On the plane $X-Z$ the region of phase-space occupied by these orbits is filled with resonant and unstable orbits. As resonant orbits, loops are also centrophobic and do not approach the galaxy centre.

\item {\bf Unstable orbits}: Unstable orbits have only two constants of motion in a three-dimensional space. As a result, they do not show a definite structure in phase-space (see the orbit $\eta=1$ on the upper panel of Fig.~\ref{fig:phase}). In contrast to box, loop and resonant orbits, systems moving on unstable orbits will quickly move off the symmetry planes\footnote{For a recent study of orbit instability in a triaxial potential such as the one considered here see Adams et al. (2007) and references therein.}. These orbits can be found more frequently around the minor and intermediate axis and do not exist on the plane formed by the major and intermediate axis (e.g. Goodman \& Schwarzschild 1981).

\end{itemize}

We emphasize that structure of the surface of section strongly depends on the apocentric radius of the orbit $r_a\approx r_0$. As Gerhard \& Binney (1985) show, the fraction of loop orbits tends to increase at large distances, whereas resonant orbits dominate the phase space volume close to the galaxy centre.

The upper and lower panels of Fig.~\ref{fig:orbXY} show the projection of the orbits examined above on the orbital planes $X-Y$ and $X-Z$, respectively. These Figures illustrate the richness of the orbit zoo allowed by a triaxial potential, even if this is just a small sample of all the possible orbital configurations. 

A clear transition between box and loop orbits can be seen in the upper panel of Fig.~\ref{fig:orbXY} as we increase the value of the orbital circularity. Orbits with $\eta\ge 0.6$ are loops, whereas orbits with $\eta=0.1,0.2, 0.4$ and 0.5 are boxes and boxlets. The only example of a resonant orbit on this plane occurs for $\eta=0.3$, which corresponds to a resonance of the type $(l,m,n)=(-5,4,0)$. The number of resonances increases on the plane $X-Z$, as we can see in the lower panel of Fig.~\ref{fig:orbXY}. For example, $\eta=0.3$ corresponds to a $(-5,0,4)$ resonance whereas $\eta=0.6$ is a''fish'' orbit with a frequency ratio of $(3,0,-2)$. Our potential allows for a much larger number of resonances, many of which appear for orbits that are not initially confined on the symmetry planes (see Merrit \& Valluri 1999 for details).

\begin{figure}
\includegraphics[width=84mm]{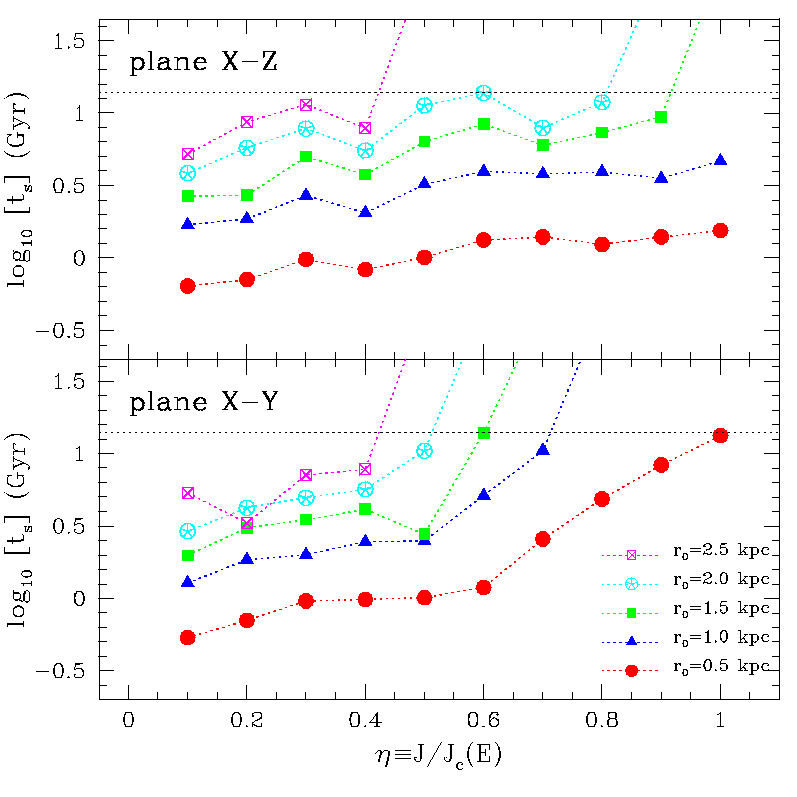}
\caption{Survival time ($t_s$) of the cluster F1 as a function of the initial distance $r_0$ to the galaxy centre and the orbital circularity $\eta$. Horizontal dotted lines indicate the value of a Hubble time $t_H=14$ Gyr. The NFW potential has a peak velocity $V_{\rm max}=20$ km/s, similar to the value estimated for the Fornax dSph.} 
\label{fig:surv}
\end{figure}

\subsection{Cluster survival time}\label{sec:surv}
For a cluster with a fixed density profile, two factors determine the amount of tidal mass loss during a Hubble time: (i) the pericentric distance of its orbit and (ii) the number of pericentric interactions. The orbital family impacts both factors, determining how often and how close a system can get to the galaxy centre. A cluster moving on a box orbit passes repeatedly through the centre of the galaxy, whereas a loop orbit always avoids the inner-most regions of the potential. As Fig.~\ref{fig:phase} shows, resonant orbits represent an intermediate case. We therefore expect that clusters on box orbits will suffer the strongest episodes of mass loss while those on loop orbits will have the largest survival times. On the other hand, the number of pericentre passages within a Hubble time depends on the orbital apocentre. Clusters with a large $r_0$ spend a large fraction of the orbital time at large distances from the galaxy centre and thus suffer a small number of pericentric interactions. As a result, we expect that the mass loss rate will decrease on average as $r_0$ increases.

We define the cluster survival time, $t_s$, as the time required for a cluster to lose 95\% of its initial mass. Although arbitrary, this definition provides an accurate estimate of the survival time of cored clusters. 
Pe\~narrubia et al. (2002) and Zhao (2004) find that cored stellar systems that are subject to tidal mass stripping show a characteristic mass evolution that can be divided into two different regimes: (i) a prolonged and slow mass loss regime that is typically followed by (ii) a sharp tidal disruption that merely spans a few crossing times. This implies that our particular definition of ``survival time'' has an inaccuracy of the order of the cluster's dynamical time, i.e. a small fraction of a Hubble time for the cluster models considered here.

We must note that, since relaxation processes speed up the rate of tidal mass stripping (see \S~\ref{sec:rel} for details), the survival times derived here represent {\it upper} values. On the other hand, the fact that our potential is kept fixed during the whole cluster evolution tends to {\it underestimate} the survival time of GCs in dwarf galaxies because, in self-consisent systems, the inner regions of dark matter haloes are expected to react when the enclosed halo mass compares with the mass of the cluster, which in practice decreases the strength of the dwarf's tidal field. 

Bearing these limitations in mind, Fig.~\ref{fig:surv} shows the survival times computed for the cluster F1 in a potential scaled to the parameters of the Fornax dSph. We have explored a wide range of apocentric distances $0\le r_{0}\le 2.5$ kpc, and note that the stellar truncation radius of Fornax occurs at $R_{t,{\rm For}}=2.1$ kpc. As the Figure shows, the survival time increases monotonically as the apocentric distance increases. 

Survival time also tends to increase with orbital circularity, although the variation is not entirely monotonic. The oscillations of $t_s=t_s(\eta)$ for a fixed value of $r_0$ are visible in both panels and arise from changes of the orbital family. For example, inspection of the systems with $r_0=2.0$ kpc in the upper panel reveals that (i) values $\eta=0.1, 0.4, 0.7$ correspond to box orbits and lead to local minima of $t_s$, as one would expect, (ii) values $\eta=0.2, 0.3, 0.5$ and 0.6 correspond to resonant orbits, and (iii) $\eta\ge 0.8$ are irregular box orbits. The results are similar for other values of $r_0$, although irregular orbits with high circularity become more common as the orbital apocentre increases.

The strong sensitivity of $t_s$ to the orbital family is also evident in the lower panel of this Figure. For example, the sharp increase in $t_s$ for $\eta\ge 0.6$, independent of the value of $r_0$, corresponds to the transition between box and loop orbits (see Fig.~\ref{fig:phase}). The latter maximize the survival times of the cluster and Fig.~\ref{fig:surv} shows that F1 does not disrupt if the loop orbit has $r_a\simgreat 1.5$ kpc. We expect therefore that the cluster debris associated with this orbital family is deposited predominately in the inner regions of dwarf galaxies (see \S~\ref{sec:sb}).

We note that a large fraction of box and resonant orbits ($\eta\le 0.4$) on both orbital planes leads to the tidal disruption of the cluster F1 within a Hubble time for the whole range of orbital apocentres explored here. Some of these orbits bring globular clusters beyond the stellar truncation radius of Fornax which, as we will see below, may (possitively) impact our ability to detect the debris.

\begin{figure*}
\includegraphics[width=130mm, height=130mm ]{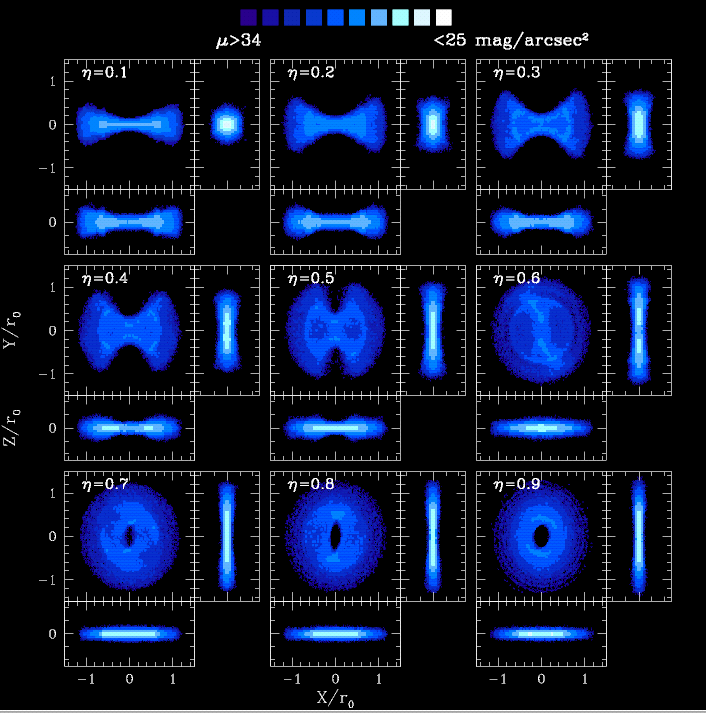}
\caption{Projected surface brightness associated with the tidal debris from the cluster F1 on different orbits initially confined on the symmetry plane $X-Y$. The initial apocentric distance is $r_0=0.5$ kpc and the pixel resolution is $0.04r_0$. Note that the distribution of debris reflects the progenitor's orbit shown in the upper panel of Fig.~\ref{fig:orbXY}.} 
\label{fig:SBXY}
\end{figure*}

\begin{figure*}
\includegraphics[width=130mm, height=130mm ]{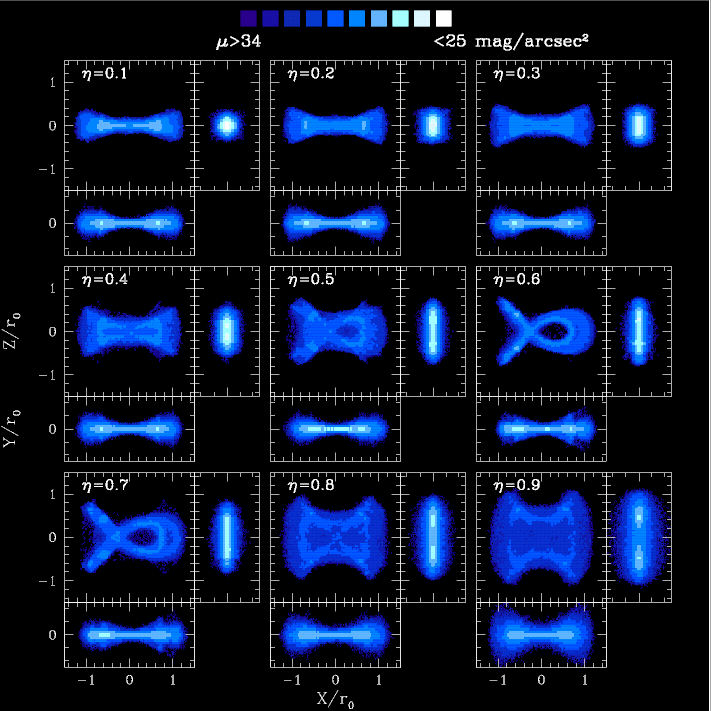}
\caption{As Fig.~\ref{fig:SBXY} for orbits initially on the $X-Z$ plane .} 
\label{fig:SBXZ}
\end{figure*}

\begin{figure}
\includegraphics[width=84mm]{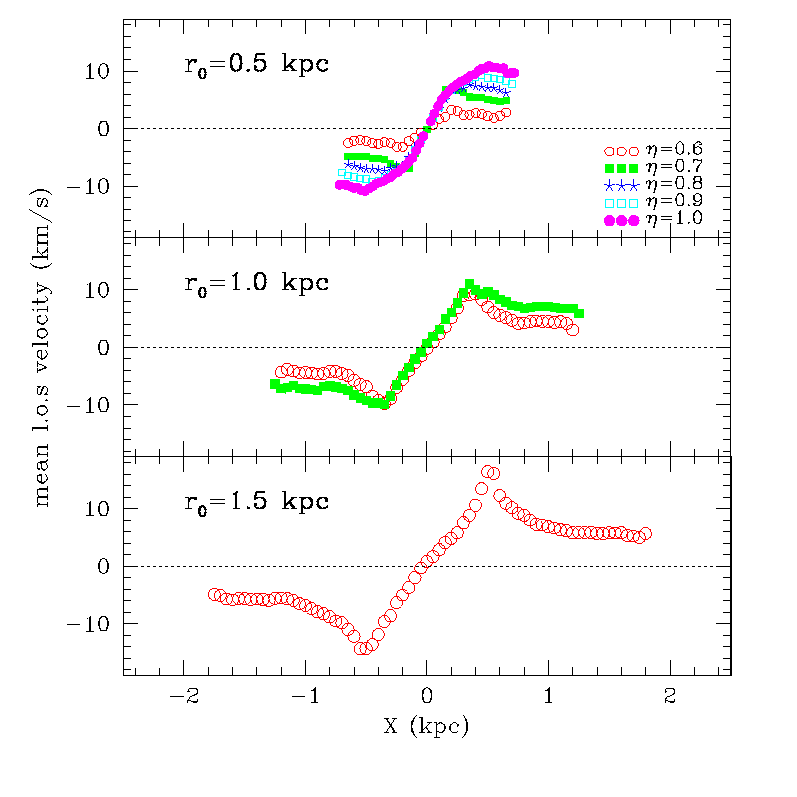}
\caption{Mean line-of-sight velocity of the cluster F1 debris. Only models moving on loop orbits on the X-Y plane are considered. The line-of-sight projection is aligned with the Y-axis to maximize the rotational velocity component. Note that, although the rotation signal strengthens as the orbital pericentre increases, the number of loop orbits that lead to the disruption of F1 decreases. We find that maximum rotational velocity can reach values of the order of the dwarf's peak velocity ($V_{\rm max}=20.6$ km/s). At large radii, however, the mean rotational velocity falls to $\langle v_{l.o.s}\rangle \simeq 5$ km/s. } 
\label{fig:rot}
\end{figure}

\subsection{Debris morphology}\label{sec:morph}
Stars that escape from a system undergoing tidal disruption have energies and angular momenta that are similar to those of the progenitor system (e.g. Pe\~narrubia et al. 2006). The scatter in these quantities causes a progressive phase-mix of debris along the progenitor's orbit until, eventually, the full phase-space volume is filled (Johnston et al. 1998, Helmi \& White 1999). 
The time that tidally stripped stars need to fill the phase space volume of the progenitor's orbit is of the order of the dwarf's crossing time.
The fact that typical dwarf galaxies have crossing times that are considerably shorter than a Hubble time (for example in the case of Fornax, which has a crossing time $t_{\rm cr}\simeq 230$ Myr, a GC may perform of the order of $t_H/t_{\rm cr}\simeq 60$ revolutions) suggests that the debris from tidally disrupted GCs fully phase-space mix on a relatively short time scale. Also, any transient feature is necessarily short-lived and disappears within few crossing times.

Because regular orbits conserve three integrals of motion in a static three-dimensional potential, they distribute in toroidal surfaces in a six-dimensional (6D) phase space (e.g. Binney \& Tremaine 2008). Unfortunately, 6D orbital information is barren to us, as only projected quantities can be known with the present instrumentation. As the {\it catastrophe} theory explains (e.g. Arnold 1992), 
the projection of a six-dimensional manifold onto a two-dimensional space necessarily results in the formation of {\it folds} and {\it cusps} (see e.g. Cappellari et al. 2004 for a schematic representation). Folds are curves on the plane that arise as two inverse images of an orbit merge and disappear, whereas cusps appear at the intersection of two or more folds. Both folds and cusps will appear as prominent features in the projection of cluster debris on the sky.

In Fig.~\ref{fig:SBXY} and~\ref{fig:SBXZ} we show the configurations of tidal debris that result from fully disrupted F1 models in a triaxial potential. The orbits of the progenitor systems are those shown in Fig.~\ref{fig:orbXY}. Note that this selection only corresponds to a small subset of the whole orbital family spectrum allowed by the potential of eq.~\ref{eq:potnfw}, and that other resonant families yield fairly different debris morphologies. The orientation of the line-of-sight direction with respect to the progenitor's orbital plane also alters the debris morphology. For simplicity, we only show here the projection along the three main orbital axes. Values of the projected surface brightness are colour-coded for clarity, and in this Figure bright areas correspond to regions densely populated by debris material in a particular projection.

These Figures show that the cluster debris neatly traces the orbits plotted in Fig.~\ref{fig:orbXY} and that, as a result, the projected morphologies are extremely rich. Folds and cusps appear in practically all orbital configuration and may produce striking features. 
For example, when looking from a direction perpendicular to the orbital plane, folds associated with box orbits exhibit a ``butterfly''-like shape, whereas loops tend to generate disc-like configurations of debris with a noticeable scarcity of stars in the inner-most regions. 
Complex stellar substructures may arise from the disruption of clusters on resonant orbits. A good example is the ``fish'' orbit shown by Fig.~\ref{fig:SBXZ}. Note, however, the strong dependence of debris morphology on the line-of-sight projection: (i) its complexity decreases if the projection angle is not aligned with the axis perpendicular to the orbital plane and (ii) the surface brightness and the area occupied by tidal debris decrease if the projection angle is aligned with the main orbital axis, which in this particular coordinate system corresponds to the X axis.

Our models show that stellar substructures arise naturally from the disruption of GCs over a Hubble time in a static potential. 
Shells (folds) and isolated over-densities (cusps) arise in almost all orbital configurations and line-of-sight projections. These features result from the triaxial shape of the galaxy potential and would not form in spherical haloes. 
Their surface brightnesses range from 36 to 26 mag/arcsec$^2$ for a progenitor cluster with an initial luminosity of $L\simeq 10^4L_\odot$. Note, however, that also clusters with {\it higher} luminosities can be fully disrupted in the tidal field of a dwarf if their initial mean density is $\langle \rho_\star\rangle \simless 0.1 L_\odot/pc^3$ (see \S\ref{sec:tidal}) and that these systems may leave behind stellar debris with surface brightneses below 26 mag/arcsec$^2$. 
 Although the latter value is around six magnitudes
brighter than the typical detection threshold that present instrumentation can reach for a resolved stellar population, the fact that they might be surrounded by a sea of field stars may hinder or even completely prevent their detection in photometric surveys. We examine this issue in more detail in \S~\ref{sec:sb}.

\begin{figure*}
\includegraphics[width=130mm, height=130mm]{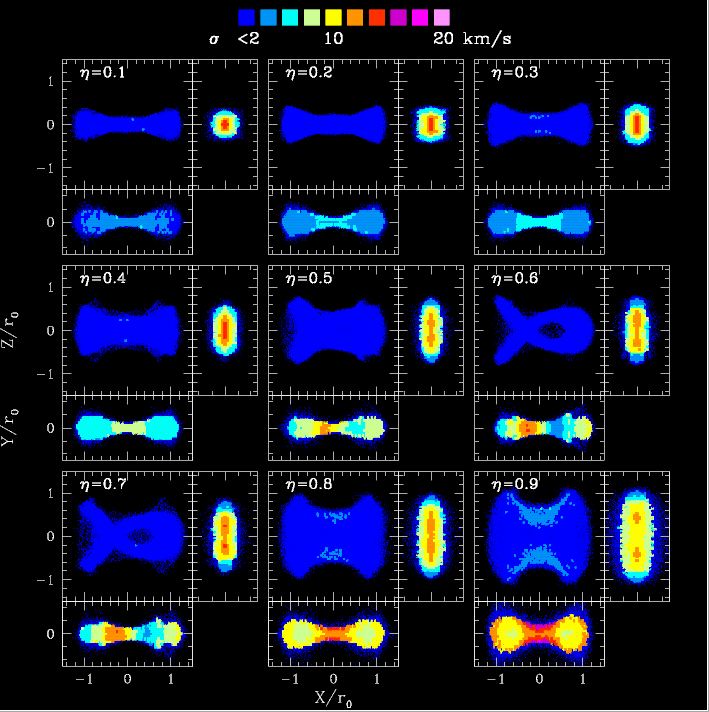}
\caption{Projected velocity dispersion of the models shown in Fig.~\ref{fig:SBXZ}. Note that all models shwon here move on either {\it box} or {\it resonant} orbits. For comparison, Fornax's field stars have a projected velocity dispersion of $\simeq 10.5$ km/s practically constant at all radii (Walker et al. 2007). Note the strong dependence on the line-of-sight projection: debris appear fairly {\it cold} ($\sigma_{l.o.s} < 3$ km/s) if the line of sight is perpendicular to the orbital plane. However, if the line of sight is aligned with the orbital plane, strong velocity dispersion gradients arise in almost all orbital configurations. 
In localized areas the tidal debris field may appear considerably {\it hotter} than Fornax's field stars, with a projected velocity dispersion that may reach $\sigma_{l.o.s} \simeq 14$ km/s. } 
\label{fig:sigXZ}
\end{figure*}

\subsection{Kinematics}\label{sec:kine}
The disruption of a GC in a dwarf galaxy also leaves kinematic imprints. The kinematic properties of the debris are fairly sensitive to the orbit of the progenitor cluster. For example, box and resonant orbits have zero net angular momentum, causing the cluster debris to have an average line-of-sight velocity that is null everywhere regardless of viewing angle.  However, loop orbits have a net rotational velocity component that will be reflected by the kinematics of the tidal debris unless the viewing angle is perpendicular to the orbital plane. 

For example, in Fig.~\ref{fig:rot} we plot the variation of the mean line-of-sight velocity $\langle v_{l.o.s}\rangle$ for a projection angle parallel to the orbital plane. This Figure includes all our F1 models on loop orbits that tidally disrupt within a Hubble time.  Note that no loop orbits lead to the full disruption of the cluster F1 for orbital apocentres $r_0\simgreat 1.5$ kpc (see models with circularity $\eta\ge 0.6$ in the lower panel of Fig.~\ref{fig:surv}). Models with large apocentres tend to exhibit a high line-of-sight peak velocity that can reach values comparable to the halo's peak velocity, $V_{\rm max, For}=20.6$ km/s. In all the models examined here the maximum of $\langle v_{l.o.s}\rangle$ occurs at $\simeq 0.3$--$0.5$ kpc from the dwarf's centre. Beyond that radius this quantity rapidly decreases, leveling off at $\sim 5$ km/s.

Although loop orbits are interesting due to the rotational component that they introduce in non-rotating systems like dSphs, the dominant orbital families in the inner regions of triaxial NFW dark matter halos are box and resonant orbits (see Fig.~\ref{fig:phase}). To examine the debris kinematics associated with these types of orbits we inspect orbits on the X-Z plane.

In Fig.~\ref{fig:sigXZ} we plot the projected velocity dispersion defined as $\sigma_{l.o.s}^2\equiv \langle (v_{l.o.s} -\langle v_{l.o.s} \rangle)^2 \rangle$ for the models shown in Fig.~\ref{fig:SBXZ}. For comparison purposes, we note that the field stars of Fornax show a flat velocity dispersion profile $\sigma_{\rm For}\simeq 10.5$ km/s (Walker et al. 2006a, 2007).  

Fig.~\ref{fig:sigXZ} displays  a strong dependence of velocity dispersion on the projection angle: cluster debris is fairly cold, $\sigma_{l.o.s}\simless 3$ km/s for all orbital circularities, so long as the projection angle is perpendicular to the orbital plane. This reflects the low velocity dispersion of the progenitor cluster F1 $\sigma_0=[4\pi G/9 R_c^2\rho_0]^{1/2}\simeq 1.34$ km/s (see Table~\ref{tab:obs}). 
In this particular line-of-sight projection, all the bright features visible in the X-Z panels of Fig.~\ref{fig:SBXZ} and~\ref{fig:sigXZ} have a velocity dispersion comparable to that of the progenitor cluster and {\it always} exhibit a high degree of symmetry with respect to the dwarf centre regardless of the progenitor's orbital family. 

Contrary to naive expectations, owing to the high radial anisotropy of these orbits the cluster debris can also appear {\it hotter} than the surrounding field stars if the line-of-sight direction lies on the orbital plane. As Fig.~\ref{fig:sigXZ} illustrates, cluster debris may have projected velocity dispersions with strong gradients throughout the galaxy and a fairly complex geometry. Box orbits tend to show a decreasing dispersion profile from the dwarf's centre, whereas off-centre, isolated hot clumps arise from resonant orbits.  
In these models, there is a clear correspondence between the brightest areas shown in Fig.~\ref{fig:SBXZ} and the ``hottest'' regions visible in  Fig.~\ref{fig:sigXZ}, which show projected velocity dispersions that can be as high as $\sim 14$ km/s.  

\section{Discussion}
\subsection{Effects of halo triaxiality}\label{sec:triaxeff}
As previously discussed, a fundamental prediction of the CDM paradigm is that dark matter haloes have a triaxial shape. There are few contributions in the literature that study the disruption of stellar systems in triaxial potentials, and it is interesting to analyze here, though briefly, how the adoption of a triaxial halo may have influenced our results.

In contrast to spherical haloes, which allow only {\it rosette} orbits, triaxial haloes allow a wide range of orbital families with very different properties, as outlined in \S\ref{sec:orbits}.
In the context of the survival of stellar clusters in dwarf galaxies, this has strong consequences. 

As loop orbits, rosettes are {\it centrophobic} and avoid the centre of the potential, with the single exception of a radial ($\eta=0$) orbit. Further, the pericentric distance and the orbital period are increasing functions of $\eta$, which implies that, at a fixed apocentre, the survival time of star clusters $t_s=t_s(\eta)$ increases monotonically.
As we discuss in \S\ref{sec:surv} this characteristic is not shared by box and resonant orbits, which can come arbitrarily close to the centre of the host in repeated occasions with independence of the value of $\eta$. As a result, clusters that move on box and resonant orbits have shorter survival times than those on rosette orbits for a given orbital circularity and apocentric distance $r_0$.

\begin{figure}
\includegraphics[width=84mm]{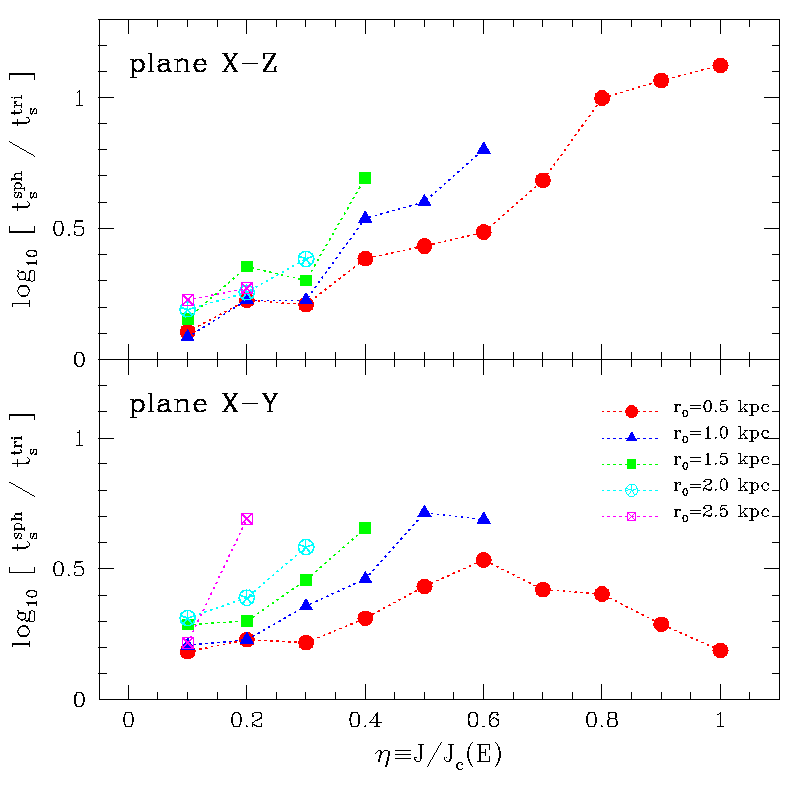}
\caption{Ratio between the survival time of the cluster F1 in a spherical ($t_s^{\rm sph}$) and a triaxial ($t_s^{\rm tri}$) halo with axis-ratios $(a,b,c)=(1,47, 1.22, 0.74)$. This quantity is plotted as a function of the initial distance $r_0$ to the galaxy centre and the orbital circularity $\eta$. The NFW potential has a peak velocity $V_{\rm max}=20$ km/s, similar to the value estimated for the Fornax dSph.} 
\label{fig:surv_sphtri}
\end{figure}

To illustrate quantitatively the change in $t_s$ introduced by the halo triaxiality, we plot in Fig.~\ref{fig:surv_sphtri} the ratio $t_s^{\rm sph}/t_s^{\rm tri}$ of the cluster F1 as a function of circularity for different orbital apocentres. This Figure shows that, independently of the orbital parameters, triaxiality always strengthens the tidal disruption of stellar clusters, so that $t_s^{\rm sph}\simgreat t_s^{\rm tri}$. The difference is smallest for the lowest values of $\eta$ (i.e. highly radial orbits in spherical potentials) and, due to the similitude between loop and rosette orbits, for the combinations of $r_0$ and $\eta$ that correspond to loop orbits in a triaxial potential (e.g. $r_0=0.5$ kpc and $\eta \geq 0.6$ in the lower panel). 
As expected, clusters on box and resonant orbits survive shorter times than on rosettes. This result is particularly evident in the upper panel of Fig.~\ref{fig:surv_sphtri}, recalling that in our triaxial halo model all orbits moving on the $X-Z$ plane are either box or resonant orbits. The difference in survival time becomes stronger as we move to high $\eta$, which is equivalent to increasing the orbital pericentre in a spherical potential, and can be as large as $t_s^{\rm sph}\approx 10 t_s^{\rm tri}$ for $\eta\simeq 1$ (i.e. circular orbits in a spherical potential).

Loop orbits and rosettes share a good number of characteristics. For example, not only do both avoid the centre of the potential, also both have a definite sense of motion around the galaxy centre. The spatial distribution of tidal debris from disrupted clusters moving on rosettes will be therefore similar to that shown in Fig.~\ref{fig:SBXY} for the models with $\eta \geq 0.6$. These tidal debris will also have a mean rotational velocity that will extend approximately out to the orbital apocentre of the progenitor cluster, in a similar fashion as discussed in \S\ref{sec:kine}. 
Finally, a noteworthy corollary is that the stellar features associated with the tidal disruption of a stellar cluster moving on box and resonant orbits (e.g. shells and isolated clumps, see Fig.~\ref{fig:SBXY} and ~\ref{fig:SBXZ}) do {\it not} arise in a spherical potential and may be used therefore to address the shape of dark matter haloes.

\begin{figure*}
\includegraphics[width=150mm, height=150mm]{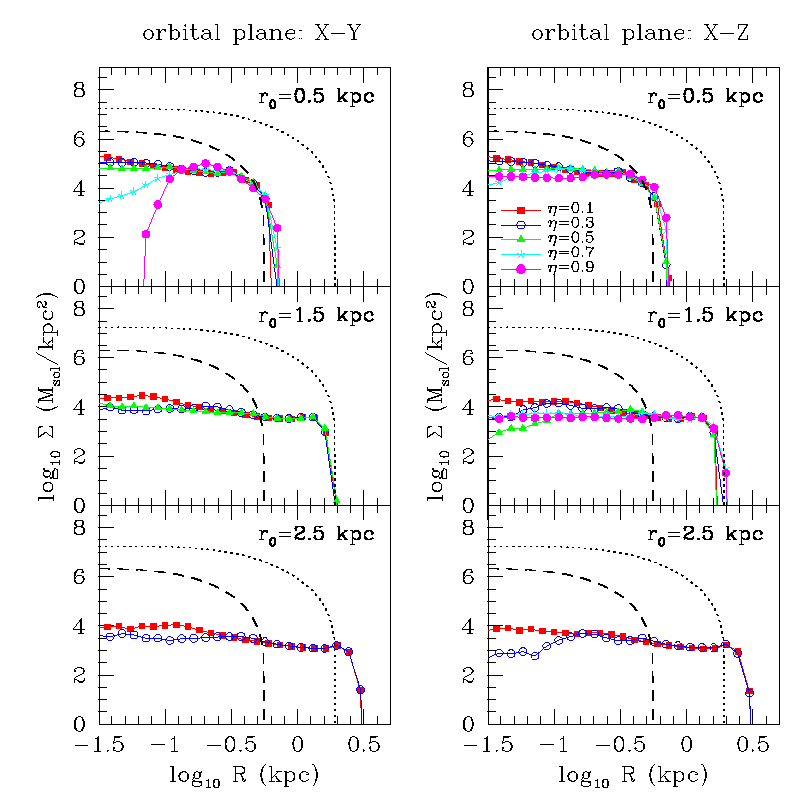}
\caption{Surface density profiles averaged on circular annuli of the cluster F1 debris. The left and right panels include models that lead to the complete tidal disruption of the cluster F1 moving on the X-Y and X-Z planes, respectively. Rows show orbits with different orbital apocentres ($r_0$). Dotted and dashed lines show the surface density profiles of Fornax and Carina, respectively. Note that in some of these models cluster debris dominates the overall light profile beyond the limiting radius of the dwarf.} 
\label{fig:sbprof}
\end{figure*}
\subsection{Detectability of Globular Cluster debris}\label{sec:sb}
In \S\ref{sec:morph} we show that the tidal disruption of a $10^4 L_\odot$ globular cluster in the triaxial potential of a typical dwarf galaxy may form stellar substructures that do {\it not} dissolve in time.  With surface brightnesses $\mu\simgreat 26$ mag/arcsec$^2$, some of these features fall within the detection limits of present instrumentation. However, most of those substructures will be surrounded by a sea of dwarf field stars, hindering the identification of cluster debris as stellar over-densities in both photometric and kinematical surveys. The detection of a disruption event will therefore be more likely in regions where the surface brightness of the cluster debris is similar to or higher than that of the field stars.

To explore this issue in some detail, Fig.~\ref{fig:sbprof} shows the surface density profiles averaged on circular annuli of the cluster F1 models that result in a complete tidal disruption within a Hubble time. We distinguish here between orbits initially confined to the X-Y (left panel) and X-Z (right panel) planes. For simplicity we consider a line-of-sight projection perpendicular to the orbital plane. We also plot the King model fits of the surface density profiles of the Fornax and Carina dwarfs (Mateo 1998) with dotted and dashed lines, respectively.
 
We begin by discussing the orbits with the smallest orbital apocentre considered in this paper, $r_0=0.5$ kpc. As we show in Fig.~\ref{fig:surv} these orbits result in the complete tidal disruption of F1 regardless of its orbital circularity $\eta$. As the top-right and left panels show, the surface density profile of the tidal debris is sensitive to the orbital family: (i) box (e.g. $\eta=0.1$) and resonant (e.g. $\eta=0.3$) orbits yield debris distributions resembling King (1962) models, i.e. cored in the inner regions with a central surface density $\Sigma_0\simless 10^5 M_\odot$/kpc$^2$, and having a sharp truncation radius at large radii. (ii) Loop orbits ($\eta \ge 0.6$ in the left panel), in contrast, yield debris distributions with a clear scarcity of stars in the inner-most regions of the dwarf. In all orbital configurations cluster debris shows a truncation radius that closely correlates with the progenitor's apocentre. In particular, we find that the distribution of debris is truncated at $R\approx 1.2 r_0$ regardless of the orbital family.

As discussed in \S\ref{sec:surv}, as we increase the orbital apocentre of the cluster F1, {\it only} box and resonant orbits lead to a complete tidal disruption within a Hubble time. This is because these orbits bring the cluster F1 repeatedly to the centre of the dwarf galaxy regardless of the orbital apocentre. A crucial result for the possible detectability of the tidal debris is that disruption occurs even for orbits with apocentres that are larger than the limiting radius of the host galaxy, i.e. $r_0\simgreat R_{t,{\rm For}}=2.1$ kpc for the Fornax dwarf. As the orbital apocentre increases, the cluster debris extends over larger volumes, leading to a drop in central surface density. We find that for apocentres $r_0=1.5$ and 2.5 kpc, the central surface density of the debris lies in the range $\Sigma_0\sim 10^3$--$10^4 M_\odot$/kpc$^2$ depending on circularity and orbital orientation. 

The maximum surface density (averaged over circular annuli) associated with tidal debris from the cluster F1 is $\Sigma\sim 10^5M_\odot$/kpc$^2$. This value is considerably lower than that of both Fornax and Carina, which implies that identifying cluster stars from the background stellar population in the central regions of dwarfs will require a sizable data set and additional kinematical information.

In contrast, cluster debris may dominate the overall light distribution at large radii if the progenitor cluster was formed beyond the limiting radius of the host galaxy (see \S\ref{sec:dynev}).
Unfortunately,  Fig.~\ref{fig:sbprof} shows that even in regions with a scarce presence of dwarf stars the detection of cluster debris may be observationally challenging in extended, luminous galaxies like Fornax given the depth that the survey should reach, approximately four decades below the central surface density. 

Finally, a noteworthy remark is that sizable spectroscopic surveys may reveal the presence of cluster debris regardless of the host background surface density if the now-defunct clusters had a chemical abundance pattern different from that of most of the dSph stars. We shall return to this issue below.
\begin{figure*}
\includegraphics[width=150mm, height=70mm]{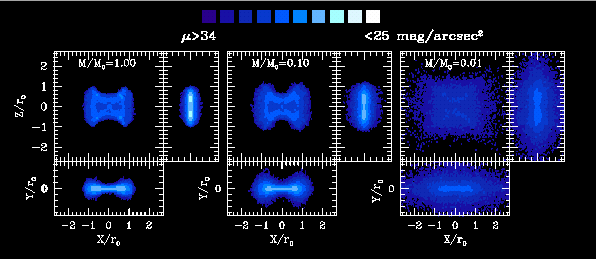}
\includegraphics[width=150mm, height=70mm]{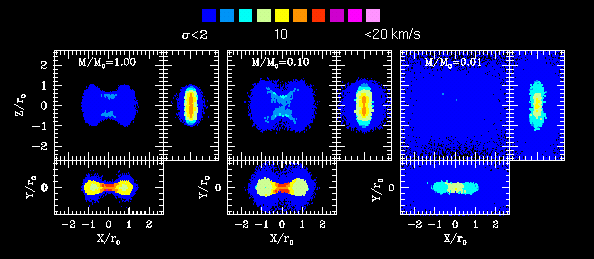}
\caption{Evolution of the spatial and kinematical distribution of cluster F1 debris in an NFW halo that loses a fraction of its initial mass to tides. This particular model corresponds to the orbit with $r_0=0.5$ kpc, $\eta=0.8$ shown in Fig.~\ref{fig:SBXZ} and~\ref{fig:sigXZ}. Note that the tidal stripping of dwarf galaxies may potentially erase stellar substructures resulting from the tidal disruption of stellar clusters.}  
\label{fig:pot_evol}
\end{figure*}

\subsection{Were Globular Clusters more common in dSphs?}\label{sec:disc}
At present, only the Sagittarius and Fornax dwarfs have associated populations of bound GCs. An understanding of the formation mechanism of these objects is still lacking, and it is unclear whether other dwarf galaxies may have also formed clusters that were subsequently tidally disrupted. It is thus interesting to carry out a {\it Gedankenexperiment} that explores whether the set of observables in the present dSph population could help us to infer if such disruption events occurred in the past.

As discussed in the previous Section, the fact that Fornax and Sagittarius are the most luminous and spatially extended dSphs of the Milky Way (Mateo 1998) may hide possible underlying stellar over-densities with low surface brightness. Fig.~\ref{fig:sbprof} shows that the potential presence of GC debris would be considerably more straightforward to detect in small, faint dwarf galaxies like Carina. According to the estimates of Pe\~narrubia et al. (2008a), Fornax and Carina are embedded in dark matter halos with similar NFW parameters, despite the fact that Fornax is a factor $\sim 100$ more luminous and has a half-light radius that is around three times larger. The similarity of their halos implies that, for a given orbit, the cluster F1 would follow practically the same dynamical evolution in both galaxies. Taking the Carina dwarf as an example, the tidal disruption of the cluster F1 would be easily detectable at approximately two decades below the central surface brightness as an excess of light beyond the King limiting radius derived from the inner profile. Interestingly, such excesses have been found in several dwarf galaxies (Carina, Majewski et al. 2005; Leo I, Sohn et al. 2007; Sculptor, Westfall et al. 2006; Ursa Minor, Mart\'inez-Delgado et al. 2001, Palma et al. 2003), although they have been often interpreted as signatures of a recent tidal interaction with the Milky Way (e.g. Mu\~noz et al. 2008 and references therein). However, Pe\~narrubia et al. (2008c) argue that the orbital motion of some of these systems (e.g. Leo I) is incompatible with the presence of unbound stellar material within the area surveyed by current data, which casts doubts on the tidal origin of the reported {\it extra-tidal} features.

The models presented in the previous Sections show that kinematic information may help to identify globular cluster debris. As discussed in \S\ref{sec:morph} and~\ref{sec:kine}, if the cluster's orbital plane is approximately perpendicular to the line-of-sight directions, shells and off-centre clumps with cold projected velocity dispersions would be a clear signature of a cluster merger. Encouragingly, the presence of stellar substructures has been reported in Fornax (Coleman et al. 2004, 2005; Olszewski et al. 2006), Canes Venatici (Ibata et al. 2006, Martin et al. 2008; also see Simon \& Geha 2007 for a discrepant result) and Sextans (Kleyna et al. 2004, Walker et al. 2006b), although some of these detections remain controversial due to their marginal statistical significance. 

The presence of an off-centre localized stellar clump, reminiscent of those detected in dwarf elliptical galaxies (e.g. Binggeli et al. 2000), is probably clearest in the Ursa Minor dwarf (Olszewski \& Aaronson 1985; Irwin \& Hatzidmitriou 1995; Wyse et al. 2002; Palma et al. 2003), and has also been confirmed by deep photometric data taken with the Wield Field Planetary Camera (WFPC2) on the HST (Battinelli \& Demers 1999). This over-density locates 13'' ($\simeq 0.021 $ core radii) from the dwarf's centre. The metalicity of the stellar clump appears indistinguishible from the bulk population of the dwarf (Eskridge \& Schweitzer 2001), which only contains a single, old stellar population (e.g., Mateo 1998; Feltzing, Gilmore \& Wyse 1999). Interestingly, Kleyna et al. (2003) also reported the presence of a {\it kinematic} substructure in UMi that locates at a projected distance of 0.23 kpc from the dwarf's centre (i.e. 1.2 core radii) and has a very low velocity dispersion, $\simeq 0.5$ km/s. The UMi substructures detected in photometric and spectroscopic data may correspond to the debris of one or more disrupted stellar clusters.
 However, it must be noted that in our triaxial halo models cold features {\it always} show a symmetric pattern with respect to the dwarf centre. The fact that only {\it a single} kinematically cold substructure has been detected may imply that a significant fraction of tidal debris remain undetected or that the origin of the cold clump is not related to a recent merger event whatsoever. A third possibility was proposed by Keyna et al. (2003), who showed that dissolved stellar clusters in spherical haloes may retain a coherent and cold spatial distribution for several orbital periods if the halo profile is cored. 
The large photometric and spectroscopic studies currently underway will be able to quantify the
presence or absence of symmetrically-distributed substructures in UMi, and
provide a clean distinction between the two possible histories: tidal
disruption in a triaxial cusped potential, and quiet evaporation in a
cored spherical potential.

Projection angles that are strongly misaligned with the normal vector of the orbital plane are more uncongenial to the identification of past mergers and lead to a large variety of possible spatial and kinematical configurations of debris that go beyond the expected ``cold stellar over-densities''. For example, if the line-of-sight direction lies close the progenitor's orbital plane, the associated debris may show strong velocity dispersion gradients (see Fig.~\ref{fig:sigXZ}). 
Also, the tidal debris from a globular cluster that moves on a loop orbit may introduce a rotational velocity component in the stellar population of the dwarf. As discussed above, a rotation signal may be straightforward to detect in dwarf galaxies, which are prototypical examples of non-rotating, pressure-supported systems, especially at radii where the overall surface brightness is dominated by cluster stars, e.g. $R\simgreat 0.6$ kpc in the Carina dwarf. Signatures, or hints, of rotation have been reported in kinematic surveys of Carina (Mu\~noz et al. 2006; Walker e tal. 2008), Leo I (Mateo et al. 2008), Sculptor (Battaglia et al. 2008; Walker et al. 2008) and Fornax (Walker et al. 2008). 
A noteworthy remark is that tidally unbound stars escaping the system through tidal tails would also produce velocity gradients along dSphs similar to those plotted in Fig.~\ref{fig:rot}. Also the transverse bulk motion of the dSph galaxy can cause small velocity gradients (Kaplinghat \& Strigari 2008; Walker et al. 2008). Further data that focus on the rotating stellar component is needed to distinguish between these scenarios.
 
In this context, chemical tagging can possibly provide the most reliable means of shedding light on the nature of the stellar over-densities and kinematic oddities detected in several dSphs. Recent measurements of the chemical abundances of different elements of the Fornax globular clusters (Letarte et al. 2006, Mottini \& Wallerstein 2008) reveal a composition that is similar to Galactic globular clusters and more metal-poor than that of the Fornax field stars. If the same result can be extrapolated to the globular clusters that may have inhabited other dwarf galaxies, a chemo-dynamical mapping of dwarf galaxies would possibly provide the most efficient method to address the origin of light excesses at large radii.

\subsection{Evolving dwarf galaxy potentials}
The results obtained in the previous Sections assume a static dSph potential throughout the evolution of our GC models. Although this approximation may be appropriate for isolated dwarf galaxies, it is unlikely to hold in detail for ``subhaloes'' orbiting in the main halo of the Milky Way. Recently, Stoehr et al. (2002), Hayashi et al. (2003), Kazantzidis et al. (2004a) and Diemand, Kuhlen \& Madau (2007) have studied the modifications suffered by NFW haloes undergoing tidal disruption within a more massive system. A recurrent result is that stripping occurs predominantly in the outer regions of the halo and reduces both $V_{\rm max}$ and $r_{\rm max}$. The reduction in $r_{\rm max}$ is larger, so that the halo concentration increases. As Hayashi et al. (2003) and Pe\~narrubia et al. (2008b) show, the variation of both parameters is solely controlled by the amount of dark matter lost to tides. Because the stellar components of most Milky Way dSphs are deeply segregated within the dark matter haloes (Pe\~narrubia et al. 2008a,b), dSphs must lose at least 90\% of their initial mass before starting to shed their stars.

To illustrate how the stripping of the outer regions of dwarf galaxy haloes
may affect existing stellar substructures, we have re-integrated the models shown in previous Sections for an additional Hubble time in dark matter haloes that have lost respectively 90\% and 99\% of their initial halo mass. 
Applying the results of Pe\~narrubia et al. (2008b), the halo peak velocity $V_{\rm max}$ drops to $\approx 0.67$ and 0.33 of its initial value and the peak velocity radius $r_{\rm max}$ becomes a factor 0.37 and 0.14 smaller, respectively. 
 
 Figure~\ref{fig:pot_evol} shows how the tidal debris of the cluster F1 respond to the potential variation. For illustrative purposes we choose an orbit that leads to the formation of a substantial amount of substructure ($r_0=0.5$ kpc, $\eta=0.8$, orbital plane X-Z) and assume that tidal mass loss does not alter the shape of the halo. 

This Figure illustrates a few interesting points. One is that the volume occupied by the cluster tidal debris expands as a result of lowering the potential depth. This in turn causes an overall decrease of the surface brightness and the prominence of the brightest over-densities wanes as a result. 
For example, the two bright clumps visible in the $Y-Z$ (upper-right) panel of the unperturbed model ($M/M_0=1$) disappear when the halo loses 90\% of its initial mass (i.e $M/M_0=0.1$). A similar effect is visible in all line-of-sight directions, which clearly indicates that the amount of stellar substructure decreases if a dwarf galaxy undergoes tidal mass stripping. 
In fact, stronger mass loss episodes may erase these stellar over-densities altogether, as the right panels ($M/M_0=0.01$) show. Cluster debris integrated in a halo that has lost 99\% of its mass appear basically as a smooth ellipsoid with no obvious substructure visible in any projection angle, despite that the halo potential is indeed triaxial.
Finally, another interesting point may be gleaned from the lower panels by noting that the tidal debris become overall ``colder''. The strong velocity dispersion gradients that are visible in the unperturbed model (left panels) tend to smooth out as kinematically distinct clumps reduce their size (see bottom-middle panel). As in the surface brightness maps, strong episodes of mass loss (i.e $M/M_0 \lesssim 0.01$) erase any previously existing kinematic substructure in the dwarf. These experiments therefore suggest that dSphs tend to become spatially and kinematically uniform as they lose their outer dark matter halo envelopes to tides.

\section{summary}
We have used N-body simulations to study the tidal disruption of Globular Clusters (GCs) in dwarf spheroidals (dSphs). We assume a cosmologically motivated scenario wherein dSphs are dark-matter dominated systems that follow an NFW profile and have a triaxial shape. GCs are modelled as N-body realizations of a King (1962) sphere. We use the Fornax and Sagittarius dwarfs, which contain five and four GCs respectively, as testbeds for analyzing the orbital and mass evolution of their GC populations under different orbital configurations. Our findings may be summarized as follows:

\begin{enumerate}
\item GCs in dSphs span over three orders of magnitude in density and two orders of magnitude in mass, in contrast with the narrow density range estimated for the inner regions of the dark matter halos in which dwarf spheroidals are embedded. For example, the value of the halo density measured at $100$ pc from the halo centre only varies a factor $\sim 3$ among all the Milky Way dwarfs. This result implies that the dynamical evolution of the Fornax clusters can be safely generalised to the rest of dwarf galaxies that contain, or used to contain, Globular Clusters.

\item Clusters in dSphs show a remarkable spatial segregation depending on their mass and density, such that the densest and most massive clusters are located at close (projected) distances from the dwarf. This may result from orbital decay induced by dynamical friction.

\item We show that the most massive clusters in Fornax and Sagittarius ($M_{GC}\simgreat 10^5 M_\odot$) have central densities ($\rho_\star(0)\simgreat 10^3 M_\odot/pc^3$) that are orders of magnitude higher than those of the underlying dark matter halos. Our N-body simulations show that these systems will survive tidal disruption in any dwarf galaxy of the Milky Way.

\item In contrast, clusters with masses $\simless 10^4M_\odot$, like the Fornax cluster F1, are much less dense ($\rho_\star(0)\simless 3 M_\odot/pc^3$) and can be fully tidally disrupted if the orbits bring them close to the dwarf centre. That introduces a large dependence between the cluster's survival time and the orbital type. At a fixed orbital apocentre, we find that {\it box, resonant} and {\it loop} orbits lead to the shortest, intermediate and longest survival times, respectively. 

\item The tidal disruption of a $10^4 L_\odot$ cluster results in the formation of stellar substructures that have surfaces brightnesses $\mu\simgreat 26$ mag/arcsec$^2$ and very complex spatial morphologies. Depending on the cluster's orbital family and line-of-sight projection, we find that cluster debris may appear as shells, isolated clumps and elongated over-densities. 

\item The kinematics of the cluster debris are strongly sensitive to the relative orientation between the line-of-sight direction and the cluster's orbital plane. On average, perpendicular and parallel orientations yield substructures that have, respectively, lower and higher projected velocity dispersions than the field stars of the host dwarf galaxy. 

\item We find that the disruption of GCs moving on loop orbits may introduce a rotational component in dSphs. Our models show that, depending on the orbital apocentre, tidal debris may reach a {\it maximum} mean rotational velocity that is comparable to the peak velocity ($V_{\rm max}$) of the host galaxy halo. 

\item A crucial remark is that these substructures arise from the potential triaxiality and do {\it not} have a transient nature. 
Indeed, the presence of distinct stellar substructures in dSphs may be a signature that gravitationally bound systems were accreted in the past. 

\item These substructures, however, may be completely erased if the host dwarf galaxy undergoes strong mass stripping.
  
\end{enumerate}

Although shells, cold stellar clumps and velocity gradients may be an indication of a past GC merger event, in general the detection of the associated debris will be challenging in areas of dSphs where the field stellar population of the host dominates the overall light profile (see \S\ref{sec:disc}). Because dwarf galaxies and GCs have distinct stellar populations, a possible remedy may be found in future chemodynamical surveys involving hundreds of stars. These data sets will likely shed light on the origin of the stellar substructures and chemodynamically distinct components detected in some Milky Way dSphs.

\vskip1cm
JP likes to thank G. Van den Ven, R. Kennicut and D. Lynden-Bell for helpful comments. We also acknowledge useful inputs from the anonymous referee.

{} 

\begin{table*}
\caption{Observational properties of the Fornax (Mateo 1998) and Sgr (Majewski et al. 2003) dSphs and their Globular Clusters (MacKey \& Gilmore 2003) }
\begin{tabular}{l c  c c c c c  } \hline \hline
{\bf Name} & Angular sep.& $[Fe/H]$ & $R_c$ & $R_t$ & $\log_{10}(L)$ & $\log_{10}[\rho_{\star}(0)]$\\
           &  (kpc)      &          & (pc)  & (pc)  & $(L_\odot)$    & $(M_\odot/pc^3)$ \\ \hline 
For dSph   &  0.00	 & $-1.3$     & $400\pm 4$ & $2078 \pm 20$& $7.13\pm 0.2$ & $-1.14 \pm 0.20$\\ \hline
F1	   &  1.60	 & $-2.25$     & $10.0\pm 0.3$ & $60 \pm 20$& $4.07\pm 0.13$     & $0.48 \pm 0.07$\\
F2	   &  1.05	 & $-1.65$     & $5.8\pm 0.2$ & $76 \pm 18$ & $4.76\pm 0.12$     & $1.78 \pm 0.07$\\
F3	   &  0.43	 & $-2.25$     & $1.6\pm 0.6$ & $63 \pm 15$ & $5.06\pm 0.12$     & $3.47 \pm 0.07$\\
F4	   &  0.24	 & $-1.65$     & $1.8\pm 0.2$ & $44 \pm 10$ & $4.69\pm 0.24$     & $3.18 \pm 0.07$\\
F5	   &  1.43	 & $-2.25$     & $1.4\pm 0.1$ & $50 \pm 12$ & $4.76\pm 0.20$     & $3.27 \pm 0.07$\\ \hline
Sgr dSph   &  0.00	 & $[-0.5, -1.3]$     & $1560\pm 20$ & $12600 \pm 20$& $7.24\pm 0.2$ & $-2.96 \pm 0.20$\\ \hline
M54	   &  0.00	 & $-1.65$     & $0.91\pm 0.04$ & $59 \pm 21$& $5.36\pm 0.08$               & $4.45 \pm 0.05$\\
Terzan 7   &  2.68	 & $-0.64$     & $1.63\pm 0.12$ & $23 \pm 8$& $3.50\pm 0.10$                & $1.97 \pm 0.07$\\
Terzan 8   &  4.40	 & $-2.25$     & $9.50\pm 0.72$ & $66 \pm 26$& $3.67\pm 0.14$               & $0.72 \pm 0.23$\\
Arp 2	   &  3.07	 & $-1.65$     & $13.67\pm 1.85$ & $139 \pm 49$& $3.59\pm 0.14$             & $0.35 \pm 0.25$\\
\hline 
\newline
\newline
\end{tabular}\label{tab:obs}
\end{table*}

\end{document}